\documentclass[journal]{IEEEtran}
\usepackage{cite}
\usepackage{times}
\usepackage{amsmath,graphicx}
\usepackage{epstopdf}
\usepackage{xcolor}
\usepackage{amssymb}
\usepackage{array}
\usepackage{booktabs}
\usepackage{multirow}
\usepackage{graphicx}
\usepackage{color}
\usepackage{float}
\usepackage{stfloats}
\usepackage[misc]{ifsym}
\usepackage{stmaryrd}
\usepackage{grffile}
\usepackage[caption=false,font=footnotesize]{subfig}
\usepackage{makecell}
\usepackage{float}
\usepackage{threeparttable}
\usepackage{enumitem}
\usepackage{setspace}
\usepackage[pagebackref=true,breaklinks=true,linkcolor=red,anchorcolor=blue, citecolor=green,colorlinks,bookmarks=false,backref=false]{hyperref}
\usepackage{tabularx}

\begin{document}

	\title{{Local Motion and Contrast Priors Driven Deep Network for Infrared Small Target Super-Resolution}}%

	\author{Xinyi~Ying, Yingqian~Wang, Longguang~Wang, Weidong~Sheng, Li~Liu\textsuperscript{*}, Zaiping~Lin\textsuperscript{*}, Shilin~Zhou

		\thanks{X.~Ying, Y.~Wang, L.~Wang, W.~Sheng, Zaiping~Lin, S.~Zhou are with the College of Electronic Science and Technology, National University of Defense Technology, P. R. China. L.~Liu is with the System Engineering, National University of Defense Technology, P. R. China. Corresponding author: L.~Liu, Zaiping~Lin. Emails: yingxinyi18@nudt.edu.cn, dreamliu2010@gmail.com, linzaiping@sina.com.
		This work was partially supported by National Key Research and Development Program of China No. 2021YFB3100800, and the National Natural Science Foundation of China under Grant 61872379. }}

	\markboth{Submitted to IEEE Transactions on Geoscience and Remote Sensing}%
	{Shell \MakeLowercase{\textit{et al.}}: Bare Demo of IEEEtran.cls for IEEE Journals}

	\maketitle

	\begin{abstract}
		Infrared small target super-resolution (SR) aims to recover reliable and detailed high-resolution image with high-contrast targets from its low-resolution counterparts. Since the infrared small target lacks color and fine structure information, it is significant to exploit the supplementary information among sequence images to enhance the target. {In this paper, we propose the first infrared small target SR method named local motion and contrast prior driven deep network (MoCoPnet) to integrate the domain knowledge of infrared small target into deep network, which can mitigate the intrinsic feature scarcity of infrared small targets. Specifically, motivated by the local motion prior in the spatio-temporal dimension, we propose a local spatio-temporal attention module to perform implicit frame alignment and incorporate the local spatio-temporal information to enhance the local features (especially for small targets). Motivated by the local contrast prior in the spatial dimension, we propose a central difference residual group to incorporate the central difference convolution into the feature extraction backbone, which can achieve center-oriented gradient-aware feature extraction to further improve the target contrast.} Extensive experiments have demonstrated that our method can recover accurate spatial dependency and improve the target contrast. {Comparative results show that MoCoPnet can outperform the state-of-the-art video SR and single image SR methods in terms of both SR performance and target enhancement. Based on the SR results, we further investigate the influence of SR on infrared small target detection and the experimental results demonstrate that MoCoPnet promotes the detection performance. The code is available at \url{https://github.com/XinyiYing/MoCoPnet}}.
	\end{abstract}
%In this paper, we explore the sparsity in image SR to improve inference efficiency of SR networks. Specifically, we develop a Sparse Mask SR (SMSR) network to learn sparse masks to prune redundant computation.


	\section{Introduction}\label{sec-intro}

	\IEEEPARstart{I}{nfrared} imaging system is all-weather in day and night and has high penetrability, sensitivity and concealment. Infrared imaging system is widely used in security monitoring, remote sensing investigation, aerospace offense-defense and other military mission. Recently, low-resolution (LR) infrared images cannot meet the high requirements of practical military mission. Therefore, it is necessary to improve the resolution of infrared images. A straightforward way to obtain high-resolution (HR) infrared images is to increase the size of infrared sensor arrays. However, due to the technical limitations of sensors and the high cost of large infrared sensor arrays, it is necessary and important to develop practical, low-cost and highly reliable infrared image super-resolution (SR) algorithms. Note that, modern autonomous driving technology requires the infrared imaging system to detect the target in a fairly long distance. Therefore, the target only occupies a very small proportion of the whole image, and is susceptible to noise and clutters. In this paper, we mainly focus on infrared small target SR task and investigate its influence on infrared small target detection.

	The special imaging mechanism and military application of infrared imaging system put forward the following requirements for infrared small target SR: {1) \textbf{High fidelity of super-resolved images.} Noise and false contours should be avoided as much as possible. 2) \textbf{High contrast of super-resolved targets.} The target contrast in the super-resolved images should be strengthened to boost the subsequent tasks. 3) \textbf{High robustness to complex scenes and noise}. Small objects are sometimes submerged in clutter and thus of low local contrast to the background. SR algorithms should be robust to various complex scenes and imaging noise. 4) \textbf{High generalization to insufficient datasets.} The lack of infrared image datasets requires that SR algorithms should achieve stable results with a relative small dataset.}

	{The motivations of our method come from data analysis, and can be summarized as:} {1) The target occupies a small proportion of the whole infrared image (generally less than 0.12\% \cite{ACM}) and lacks color and fine structure information (\textit{e.g.,} contour, shape and texture). Few information is available for SR within a single image. Therefore, we perform SR on image sequences to use the supplementary information among the temporal dimension to improve the SR performance and the target contrast. 2) Due to the long distance between the target and the imaging system, the mobility of the targets on the imaging plane is limited, leading to small motion of the target between neighborhood frames (\textit{i.e.,} local motion prior \cite{liu2020small,sun2020infrared} in spatio-temporal dimension). Therefore, we design a local spatio-temporal attention (LSTA) module to perform implicit frame alignment and exploit the supplementary information in the local spatio-temporal neighborhood to enhance the local features (especially for small targets). 3) Compared with the background clutter, the contrast and gradient between the target and the background in the local neighborhood are high in all directions (\textit{i.e.,} local contrast prior \cite{detect61,ALCNet} in spatial dimension). Therefore, we design a center difference residual group (CD-RG) to achieve center-oriented gradient-aware feature extraction, which can encode the local contrast prior to further improve the target contrast.}

	{Based on the above observations, we propose a local motion and contrast prior driven deep network (MoCoPnet) for infrared small target SR. The main contributions can be summarized as follows:
	1) We propose the first infrared small target SR method named local motion and contrast prior driven deep network (MoCoPnet) and summarize the definition and requirements of this task. The proposed modules (\textit{i.e.,} central difference residual group and local spatio-temporal attention module) of MoCoPnet integrate the domain knowledge (\textit{i.e.,} local contrast prior and local motion prior) of infrared small targets into deep networks, which can mitigate the intrinsic feature scarcity of data-driven approaches \cite{ALCNet}.
	2) The experimental results demonstrate that MoCoPnet can achieve state-of-the-art SR performance and effectively improve the target contrast.
	3) Based on the SR results, we further investigate the influence of SR on infrared small target detection. The experimental results show that MoCoPnet can promote the detection performance to achieve high signal-to-noise ratio gain (SNRG), signal-to-clutter ratio gain (SCRG), contrast gain (CG) scores and improved receiver operating characteristic curve (ROC) results.}

	\section{Related Work}\label{sec-related}

	\subsection{Single Image SR}
	Image SR is an inherently ill-posed optimization problem and has been investigated for decades. In literature, researchers have proposed a variety of classic single image SR (SISR) methods, including prediction-based methods \cite{survey12,survey10}, edge-based methods \cite{survey14,survey13}, statistics-based methods \cite{survey15,survey16}, patch-based methods \cite{survey13,survey18} and sparse representation methods \cite{survey20,survey21}. However, most of the aforementioned traditional methods use handicraft features to reconstruct HR images, which cannot formulate the complex SR process and thus limits the SR performance. Recently, due to the powerful feature representation capability, convolutional neural networks (CNNs) have been widely used in single image SR task and achieve the state-of-the-art performance \cite{RCAN,PAM}. Dong \textit{et al.} \cite{SRCNN} proposed the pioneering CNN-based work SRCNN to recover an HR image from its LR counterpart. Kim \textit{et al.} \cite{VDSR} deepened the network to 20 convolutional layers (\textit{i.e.,} VDSR) and achieved improved SR performance by increasing model complexity. {Moreover, various increasingly deep and complex architectures (\textit{e.g.,} residual networks \cite{EDSR}, recursive networks \cite{li2019lightweight,DRCN,DRRN,PR1}, densely connected networks \cite{li2020mdcn,SRDenseNet,RDN}, attention-based networks \cite{RCAN,SAN}) have also been applied to SISR for performance improvement.} Other than tackling image average distortion by norm loss, generative adversarial image SR networks \cite{SRGAN,ESRGAN} employed the perceptual loss for perceptual quality improvement.

	\subsection{Video SR}
	%Compared with SISR, video SR methods use the additional information in temporal dimension (\textit{i.e.,} temporal information) to improve SR performance.

	Existing video SR methods commonly follow a three-step pipeline, including feature extraction, motion compensation and reconstruction \cite{D3Dnet}. Traditional video SR methods \cite{videoSR-t1,videoSR-t2} employ handcrafted models to estimate motion, noise and blur kernel and reconstruct HR video sequences. Recent deep learning-based video SR methods are better in exploiting spatio-temporal information by its powerful feature representation capability and can achieve the state-of-the-art performance. Liao \textit{et al.} \cite{video-m1} proposed the pioneering CNN-based video SR method to perform motion compensation by optical flow and then ensembled the compensated drafts via CNN. Afterwards, A series of optical flow-based video SR algorithms \cite{VESPCN,SOFVSR20} emerged to explicitly perform motion estimation and frame alignment, resulting in vague and duplication \cite{vague2}. To avoid the aforementioned problem, deformable convolution \cite{DCN1,DCN2} has been employed to perform motion compensation explicitly in a unified step \cite{TDAN,EDVR} through extra offsets. Apart from these explicit motion compensation methods, implicit approaches (\textit{e.g.,} 3D convolution networks \cite{DUF-VSR,FSTRN}, recursive networks \cite{over-video30,over-video35}, non-local networks \cite{EDVR,PFNL}) have also been applied to video SR for performance improvement.

	\subsection{Infrared Image SR}
	With the increased demands of high-resolution infrared images, some researchers perform image SR on infrared images. Traditional methods \cite{InMao} consider SR as sparse signal reconstruction in compressive sensing. Based on the previous studies, Zhang \textit{et al.} \cite{InZhang} combined compressive sensing and deep learning to achieve improved SR performance with low computational cost. Han \textit{et al.} \cite{InHan} proposed to employ CNNs to recover high-frequency components with upscaled LR images to generate the SR results. He \textit{et al.} \cite{InHe} proposed a cascaded deep network with multiple receptive fields for large scale factor ($\times$8) infrared image SR. Liu \textit{et al.} \cite{InLiu} proposed to use generative adversarial network and perceptual loss to reconstruct the texture details of infrared images. {Chen \textit{et al.} \cite{IERN} employed an iterative error reconstruction mechanism to perform SR in a coarse-to-fine manner. Huang \textit{et al.} \cite{PSRGAN} proposed a progressive super-resolution generative adversarial network and employed the multistage transfer learning strategy to improve the SR performance from small samples. Prajapati \textit{et al.} \cite{ChaSNet} proposed channel splitting-based convolutional neural network to eliminate the redundant features for efficient inference. Yang \textit{et al.} \cite{Visible-Assisted} proposed a  visible-assisted training strategy to promote details preservation.}

	\subsection{Attention Mechanism}
	Since the importance of each spatial location and channel is not uniform, Hu \textit{et al.} \cite{SENet} proposed SeNet for classification, which consists of selection units to control the switch of passed data. Zhang \textit{et al.} \cite{RCAN} proposed a channel attention mechanism to calculate the importance along the channel dimension for channel selection. Anwar \textit{et al.} \cite{FeatureA} proposed feature attention to urge the network to pay more attention to the high frequency region. Dai \textit{et al.} \cite{SAN} proposed second-order attention to adaptively readjust features for powerful feature correlation learning. Wang \textit{et al.} \cite{Wang2021Exploring} explored the sparsity in SR task and proposed sparse masks for efficient inference. The spatial mask and channel mask calculate the importance along both the spatial dimension and the channel dimension to prune the redundant computations. The aforementioned studies only consider the global importance on spatial and channel dimension. Since small targets only occupy a small portion in the whole image and have high contrast with the local neighborhood, we design a local attention mechanism which can better characterize the small targets.

	\subsection{Sequence Image Infrared Small Target Detection}
	Sequence image infrared small target detection is significant for long-range precision strikes, aerospace offensive-defensive countermeasures and remote sensing intelligence reconnaissance. According to whether the sequential information is used, sequence image infrared small target detection methods can be divided into two categories: detect before track (DBT) methods and track before detect (TBD) methods. {Based on the results of single image infrared small target detection \cite{IPI,detect59,ALCNet,liu2021non,DNAnet}, DBT methods employed the motion trajectory of targets through sequence image projection to eliminate the false targets and reduce the false alarm rate.} DBT methods have low computational cost and are easy to implement. However, the performance drops rapidly with low SNR. TBD methods \cite{application,Novel,improved} commonly follow a three-step pipeline, including background suppression, region of interest extraction and target detection. TBD methods are robust to images with low SNR but have high computational cost, which cannot meet the requirements of real-time detection. It is challenging to achieve high detection rate and low false alarm rate in real-time due to the lack of target information, the complex background noise, the insufficient public datasets and the explosion of data amount and the computational cost. Therefore, it is necessary to recover reliable image details and enhance the contrast between target and background for detection. 

	\begin{figure*}[t]
		\centering\includegraphics[width=17.5cm]{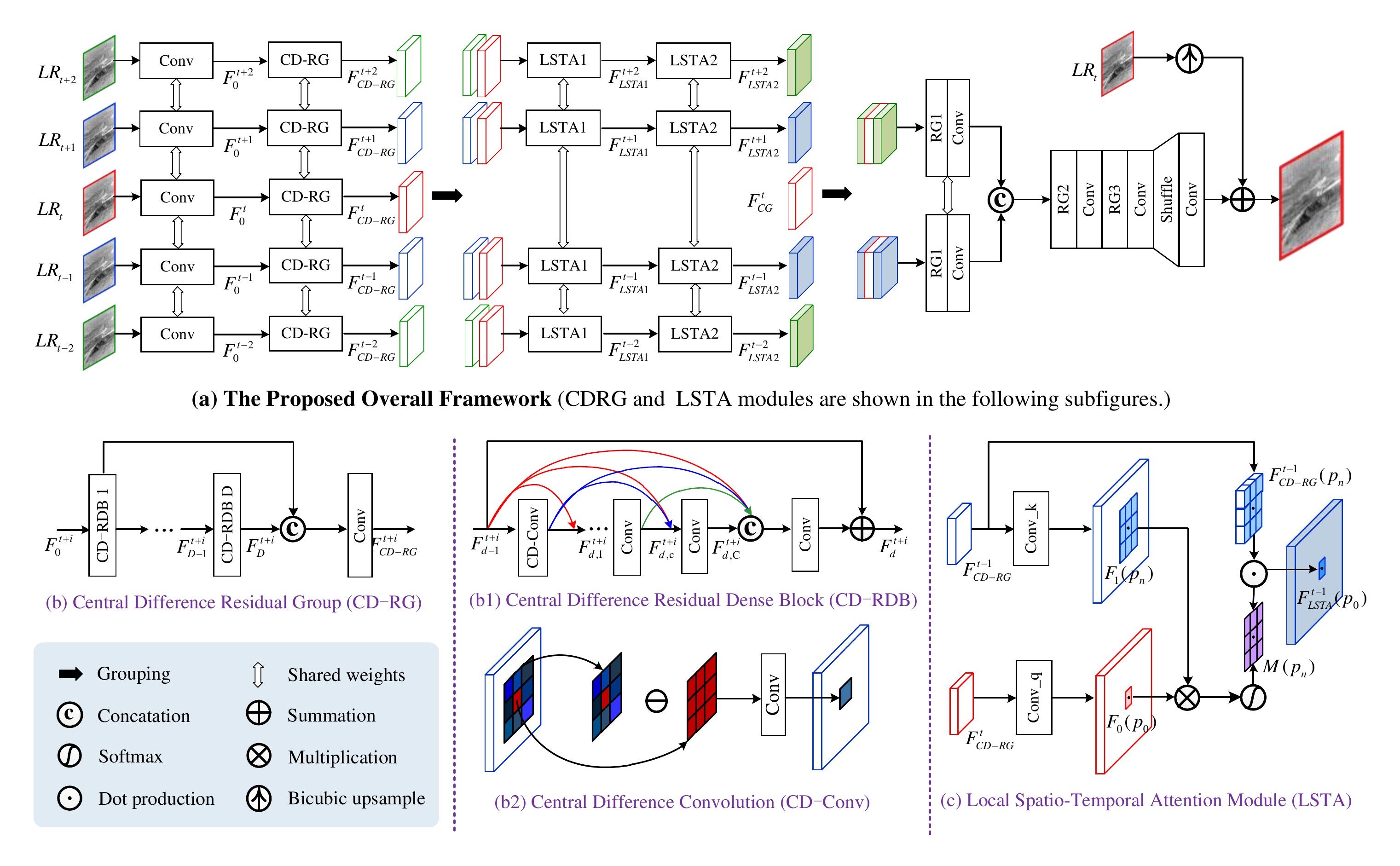}
		\caption{{The proposed architecture of MoCoPnet. (a) represents the overall framework. (b) represents the central difference residual group (CD-RG) and (b1), (b2) represents its sub-modules central difference dense block (CD-RDB) and central difference convolution (CD-Conv) respectively. (c) represents the local spatio-temporal attention (LSTA) module with kernel size 3 and dilation rate 1.}}\label{fig-Network}
	\end{figure*}

	\section{Methodology}\label{Methodology}
	In this section, we introduce our method in details. Specifically, Section \ref{Sec-net} introduces the overall framework of our network. Section \ref{Sec-RG}-\ref{Sec-LSTA} introduce the two modules which integrate local contrast prior and local motion prior of infrared small target into deep networks.
	\subsection{Overall Framework}\label{Sec-net}

	The overall framework of our MoCoPnet is shown in Fig.~\ref{fig-Network}. {Specifically, an image sequence with 5 frames $LR_{t+i}$ $(i=[-2,2])$ is first sent to a convolutional layer to generate the initial features $F_{0}^{t+i}$ $(i=[-2,2])$, which are then sent to the central difference residual group (CD-RG) to achieve center-oriented gradient-aware feature extraction. Then, each neighborhood feature $F_{CD-RG}^{t+i}$ $(i=-2,-1,1,2) $ is paired with the reference feature $F_{CD-RG}^{t}$ and sent to two local spatio-temporal attention (LSTA) modules to achieve motion compensation and enhance the local features. Next, the reference feature $F_{CD-RG}^{t}$ is concated with two compensated neighborhood frames $F_{LSTA2}^{t+k},$ $F_{LSTA2}^{t-k}$ $(k=1,2) $ and then sent to a residual group (RG) and a convolution layer for coarse fusion. Afterwards, the two fused features are concatenated and sent to an RG and a convolution for fine fusion. Then, the fused feature is processed by an RG, a sub-pixel layer and a convolutional layer for SR reconstruction and upsampling. Finally, the SR reference frame is obtained by adding the bicubicly upsampled LR reference frame to accelerate the training convergence. Note that, the number of the input frames is set to 7 in this paper and the process is the same as in Fig.~\ref{fig-Network}(a).} We use the mean square error (MSE) between the SR reference frame and the groundtruth reference frame as the loss function of our network.

\begin{figure}[t]
	\centering
	\includegraphics[width=7.5cm]{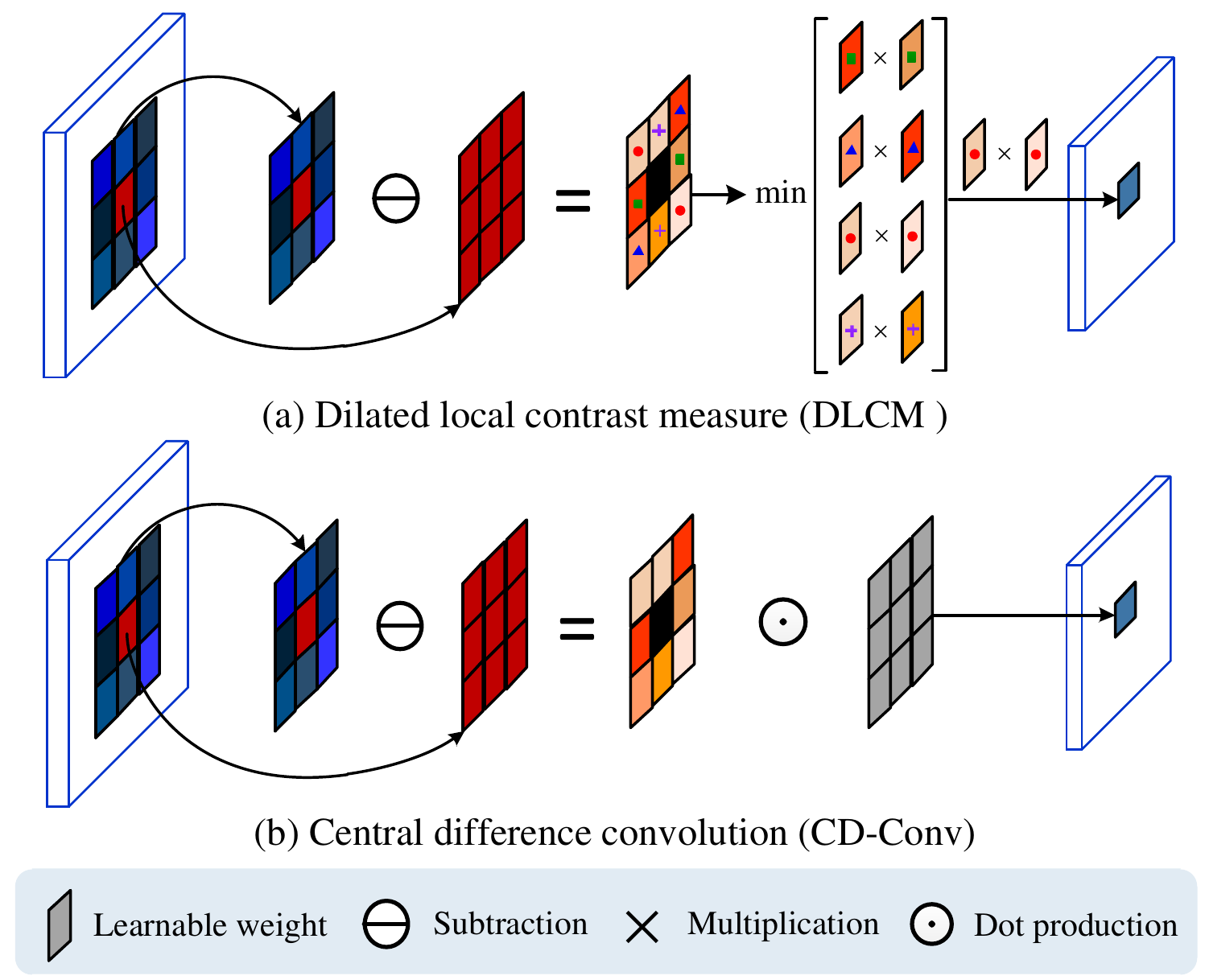}
	%	\vspace{-0.3cm}
	\caption{{Differences between (a) dilated local contrast measure (DLCM \cite{ALCNet}) and (b) central difference convolution (CD-Conv \cite{CDC,CDC2}).}} \label{fig-LCM}
\end{figure}

	\subsection{Central Difference Residual Group}\label{Sec-RG}
	{Central difference residual group (CD-RG) incorporates central difference convolution (CD-Conv \cite{CDC,CDC2}) into residual group (RG \cite{RCAN,RDN}) to achieve the center-oriented gradient-aware feature extraction, which can utilize the spatial local salient prior to strengthen the contrast of the small targets. Note that, we employ RG as the backbone of our MoCoPnet for the following reasons: RG can generate features with large receptive field and dense sampling rate, which promotes the information exploitation. The reuse of hierarchical features not only improves the SR performance \cite{Wang_2021_CVPR} but also maintains the information of small targets \cite{ACM, DNAnet,DSFNet}.}

	{The architecture of central difference residual group (CD-RG) is shown in Fig.~\ref{fig-Network}(b). The input feature $F_{0}^{t+i}$ is first fed to $D$ central difference residual dense blocks \cite{RDB} (CD-RDB) to extract hierarchical features. Then, the hierarchical features are concatenated and fed to a 1$\times$1 convolutional layer to generate output feature $F_{CD-RG}^{t+i}$. As is shown in Fig.~\ref{fig-Network}(b1), $1$ CD-Conv and $K-1$ Convs with a growth rate of $G$ are used within each CD-RDB to achieve dense feature representation. The architecture of CD-Conv is shown in Fig.~\ref{fig-Network}(b2). CD-Conv aggregates the center-oriented gradient information, which echoes the spatial local saliency prior of infrared small target. {As shown in Fig.~\ref{fig-LCM}, different from handcrafted dilated local contrast measure (DLCM \cite{ALCNet}) which can only reserve the contrast information in one direction, CD-Conv is a learnable measure and can improve the contrast of small target while maintaining the background information. In conclusion, CD-Conv is more in line with the task of infrared small target SR (\textit{i.e.,} recovering reliable and detailed high-resolution image with high-contrast target). DLCM and CD-Conv can be formulated as $f(x,y)$ and $g(x,y)$:
	\begin{equation}
		\footnotesize{
			f(x,y) = {\min _{(i,j) \in \Omega^{+} }}\left\{ {\left( {S_{x,y}-S_{x-i,y-j}} \right) \left( {S_{x,y}-S_{x+i,y+j}} \right)} \right\},
		}
	\end{equation}
	\begin{equation}
		\footnotesize{
			g(x,y) =\sum\limits_{(i,j) \in \Omega } {{\omega _{i,j}}\left( {S_{x+i,y+j}}-\theta {S_{x,y}} \right)},
		}
	\end{equation}
	where $S_{x,y}$ represents the value of a specific location $(x,y)$ in the feature map, and $(i,j) \in \Omega^{+}={(d,d),(d,0),(d,-d),(0,d)}$ is the direction index. $\omega _{i,j}$ is a learnable weight to continuously optimize the local contrast measure and $\theta \in [0, 1]$ is a hyperparameter to balance the contribution between gradient-level detailed information and intensity-level semantic information. Note that, $\theta$ is set to 0.7 \cite{CDC} in our paper.}}

\begin{figure}[t]
	\centering\includegraphics[width=9cm]{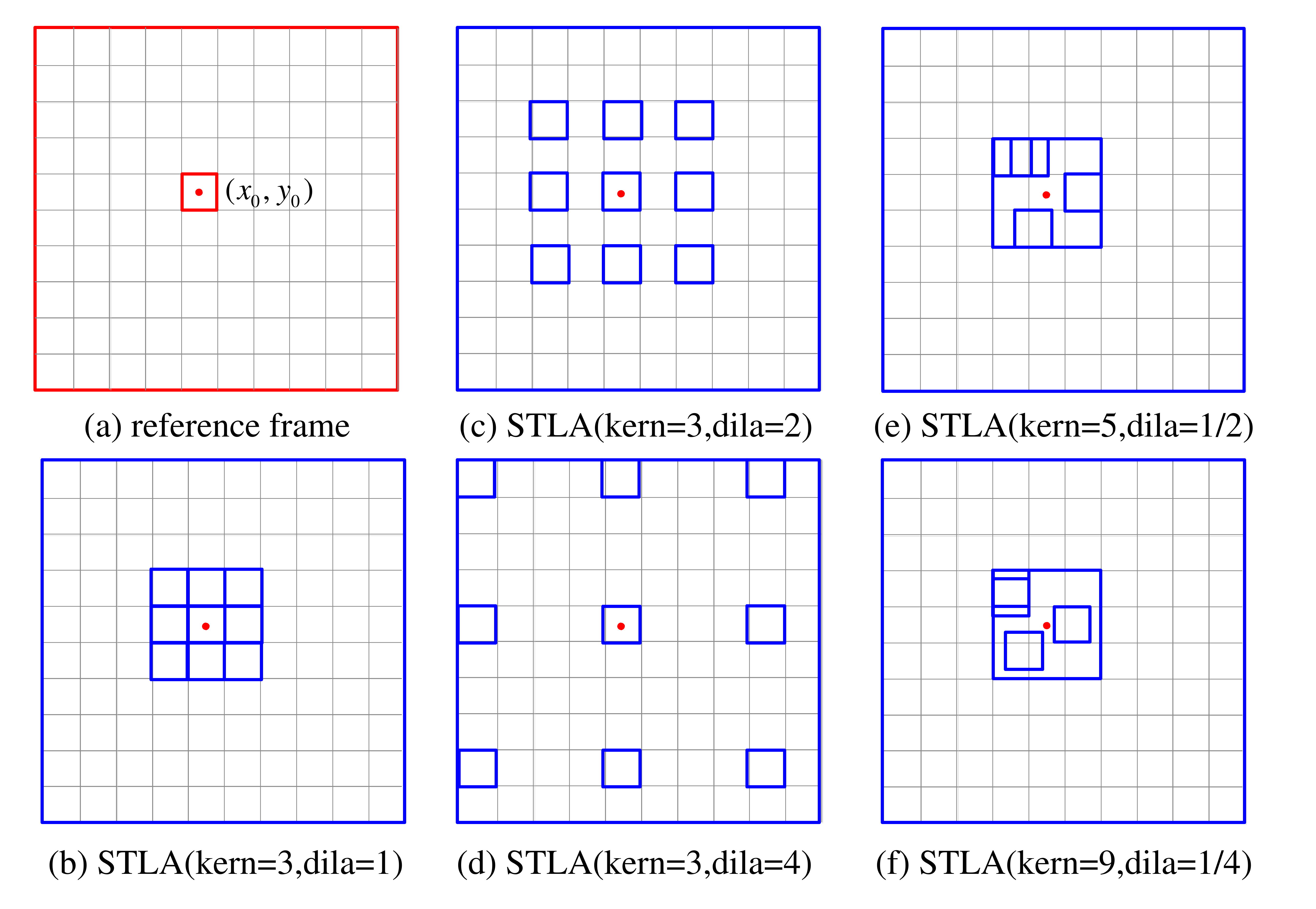}
	\caption{{An illustration of local spatio-temporal attention (LSTA) module with difference kernel size of $kern$ and dilation rate of $dila$. (a) represents the reference frame and pixel $(x_0,y_0)$ is highlighted by a red box. (b)-(f) represent the corresponding neighborhood pixels centered in $(x_0,y_0)$ and are highlighted by blue boxs.}}\label{fig-subpixel}
\end{figure}

	\subsection{Local Spatio-Temporal Attention Module}\label{Sec-LSTA}
Local spatio-temporal attention (LSTA) module calculates the local response between the neighborhood frame and the reference frame and uses the local spatio-temporal information to enhance the local features of the reference frames. The inputs of LSTA are the reference frame and one neighborhood frame. For a sequence with 7 frames, the operation need to be repeated 6 times. {The architecture of LSTA is shown in Fig.~\ref{fig-Network}(c). The red reference feature $F_{CD-RG}^{t}\in \mathbb{R}^{H\times W\times C}$ and the blue neighborhood feature $F_{CD-RG}^{t-1}\in \mathbb{R}^{H\times W\times C}$ are first fed to 1$\times$1 convolutional layers $conv\_q$ and $conv\_k$ for dimension compression to generate $F_0,F_1\in \mathbb{R}^{H\times W\times C/cr}$, where $cr$ is the compression ratio and is set to 8 in our paper. The process can be formulated as:
		\begin{equation}\label{eq-a}
			\begin{split}
				{F}_{0}&=H_{conv\_q}\left(F_{CD-RG}^{t}\right),\\
				{F}_{1}&=H_{conv\_k}\left(F_{CD-RG}^{t-1}\right),
			\end{split}
		\end{equation}
		where $H_{conv\_k}$ and $H_{conv\_q}$ represent 1$\times$1 convolutions. Then, we calculate the response between each location $p_0$ in $F_0$ and the corresponding neighborhood (centered in $p_0$) in $F_1$. Afterwards, the response is summed and softmax along the channel dimension to generate the attention map $M$.} The process is defined as:
	\begin{align}\label{eq-M}
		M(p_n) ={\rm{softmax}}(\sum_{k=1}^{C/rd}F_{0}(p_0,k) \cdot F_{1}(p_n, k)),
	\end{align}
	where $p_n$ represents the $n^{th}$ value of the local neighborhood centered in $p_0$ with kernel size of $kern$ and dilation rate of $dila$. The purple 3$\times$3 grid in Fig.~\ref{fig-Network}(c) is the local attention feature map with parameter ($kern$=3, $dila$=1). {Note that, as shown in Figs~\ref{fig-subpixel}(c) and (d), $dila$ can be integer larger than 1 to enlarge the receptive filed without additional computational cost. As shown in Figs~\ref{fig-subpixel}(e) and (f), $dila$ can also be fractional to capture the sub-pixel motion between frames and we employ bilinear interpolation to generate the exact corresponding values.}

	{Finally, dot production is performed between the local neighborhood feature $F_{CD-RG}^{t-1}(p_n)$ centered in $p_0$ and the corresponding attention map $M(p_n)$ to generate the value of location $p_0$ in the output feature $F_{LSTA}^{t-1}(p_0)$. The process is formulated as:
		\begin{align}\label{eqout}
			F_{LSTA}^{t-1}(p_0) =\sum_{\forall p_n\in G}F_{CD-RG}^{t-1}(p_n)\cdot M(p_n).
	\end{align}}

	{LSTA first calculates the response between the reference frame and its adjacent frames to generate the attention map, and then calculates a weighted summation of these frames using the generated attention maps. In this way, the neighborhood frames can be implicitly aligned and the complementary temporal information can be incorporated to enhance the features of small targets.}
	\vspace{-.5pt}

	\section{Experiments}\label{sec4}
	In this section, we first introduce the experiment settings, and then conduct ablation studies to validate our method. Next, we compare our network to several state-of-the-art SISR and video SR methods. Finally, we investigate the influence of SR on infrared small target detection.

	\subsection{Experiment Settings}
	In this subsection, we sequentially introduce the datasets, the evaluation metrics, the network parameters and the training details.

	\subsubsection{Datasets}\label{sec-Datasets}
	{Hui \textit{et al.} \cite{infrareddata} developed a dataset for detection and tracking of dim-small aircraft infrared targets under ground/air background. This dataset contains 22 image sequences (totally 16177 frames) with a resolution of 256$\times$256. Recently, a large-scale high-quality semi-synthetic dataset (named SAITD \cite{SAITD}) has been proposed for small aerial infrared targets detection. SAITD dataset contains 350 image sequences with a resolution of 640$\times$512 (175 image sequences with target annotations and 175 without, 150185 images in total). The 2nd Anti-UAV Workshop \& Challenge (Anti-UAV \cite{Anti-UAV}) releases 250 high-quality infrared video sequences with multi-scale UAV targets. In this paper, we employ the $1^{st}-50^{th}$ sequences with target annotations of SAITD as the test datasets and the remaining 300 sequences as the training datasets. In addition, we employ Hui and Anti-UAV as the test dataset to test the robustness of our MoCoPnet to real scenes. In Anti-UAV dataset, only the sequences with infrared small target \cite{ACM} (21 sequences in total) are selected as the test set. Note that, we only use the first 100 images of each sequence for test to balance computational/time cost and generalization performance.}

	\begin{figure}[t]
		\centering
		\includegraphics[width=9cm]{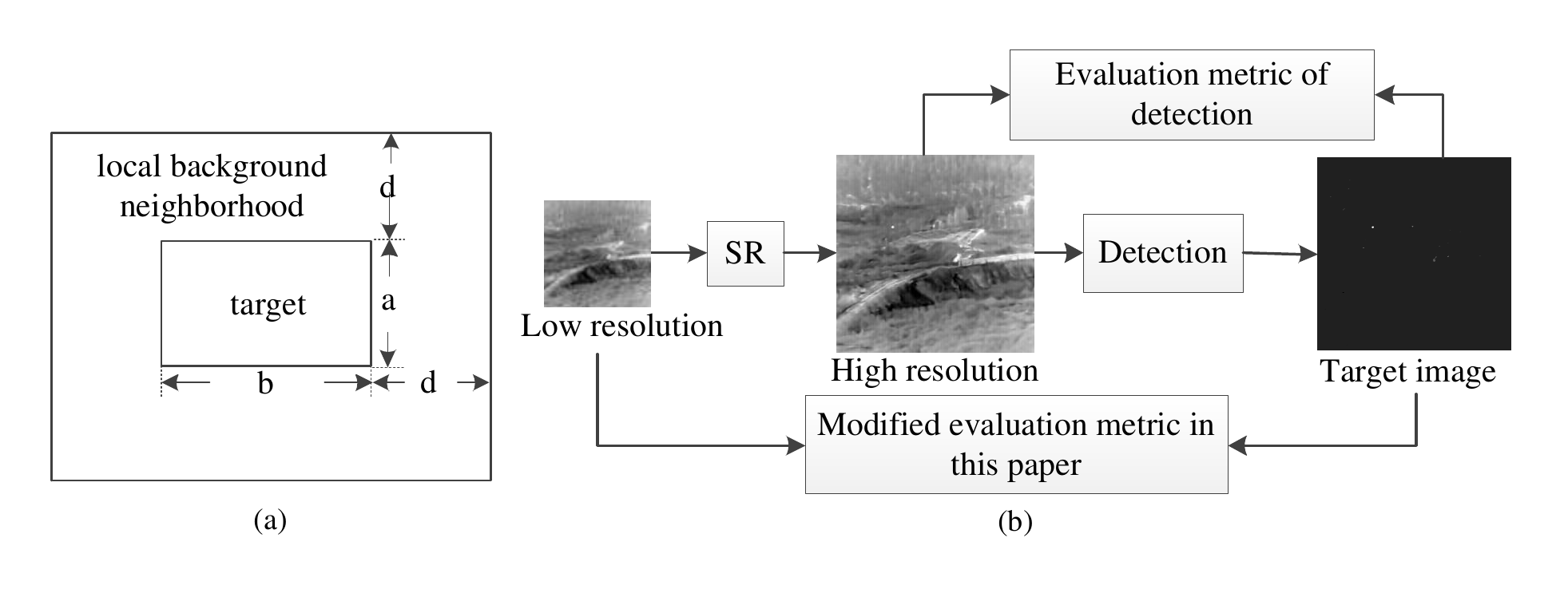}
		\caption{Evaluation metrics. (a) represents the local background neighborhood and (b) represents the modified evaluation metrics in this paper.}\label{fig_evaluation}
		\vspace{-5pt}
	\end{figure}

	\subsubsection{Evaluation Metrics}\label{sec-metric}
	We employ peak signal-to-noise ratio (PSNR) and structural similarity index (SSIM) to evaluate the SR performance. {In addition, we introduce signal-to-noise ratio (SNR) and contrast ratio (CR) in the local background neighborhood \cite{IPI} of targets to evaluate the performance of recovering small targets.} As shown in Fig.~\ref{fig_evaluation}(a), the size of the target area is $a\times b$, and the local background neighborhood is extended from the target area by $d$ both in width and height. {Note that, the parameters of local background neighborhood $(a,b,d)$ in HR images are set to $(7,7,30)$, $(11,11,50)$, $(21,21,100)$ in SAITD\footnote{{The synthetic target size in SAITD is preset to less than 7$\times$7.}}, Hui and Anti-UAV\footnote{{The target size is less than 0.12\% of the image size \cite{ACM} (\textit{i.e.,} 256$\times$256 in Hui and 640$\times$512 in Anti-UAV).}} respectively. When 4$\times$ SR is performed on HR images, the parameters (a, b, d) are set to $(29,29,120)$, $(45,45,200)$, $(85,85,400)$. When 4$\times$ downsampling is performed on HR images, the parameters are set to $(3,3,10)$, $(3,3,10)$, $(5,5,20)$.}

	To further evaluate the impact of SR algorithms on infrared small target detection, we adopt SNR gain (SNRG), background suppression factor (BSF), signal-to-clutter ratio gain (SCRG), contrast gain (CG) and receiver operating characteristic curve (ROC) for comprehensive evaluation.
	%Note that, ROC has nothing to do with the inputs of the detection algorithms, and can directly act on the detection result to describe the trend between the detection probability and the false alarm probability.
	Note that, the common detection evaluation metrics calculate the ratio of the statistics in the local background neighborhood before and after detection. Since we first super-resolve the LR image and then perform detection, the inputs of detection algorithms, which are the outputs of different SR algorithms, are different. Therefore, direct using the common detection evaluation metrics cannot evaluate the impact of SR on detection accurately. To eliminate the influence of different inputs, we modify the first four metrics to calculate the ratio of the statistics in the local background neighborhood between the LR image before SR and the HR target image after detection. The modified evaluation metrics are shown in Fig.~\ref{fig_evaluation}(b). We then introduce the aforementioned evaluation metrics in details. SNRG is used to measure the SNR improvement of detection algorithms and is formulated as:
	\begin{equation}\label{eq11}
		\begin{split}
			f_{\rm{SNRG}} &= \frac{{\rm{SNR}}^{out}}{{\rm{SNR}}^{in}}=\frac{(P_t/P_b)^{out}}{(P_t/P_b)^{in}},
		\end{split}
	\end{equation}
	where $[\cdot]^{in}$ and $[\cdot]^{out}$ represent the metrics in the local background neighborhood of the LR images and the HR target images respectively. $P_t$ and $P_b$ are the maximum value of the target area and the background area respectively. BSF is used to measure the background suppression effect and is formulated as:
	\begin{align}\label{eq12}
		f_{\rm{BSF}}= \frac{\sigma_b^{in}}{\sigma_b^{out}},
	\end{align}
	\begin{table}[t]
		\centering
		\scriptsize
		\vspace{-0.2cm}
		\renewcommand\arraystretch{1.2}
		\caption{\scriptsize{Ablation results of DLCM, Conv and CD-Conv for $4\times$SR on SAITD, Hui and Anti-UAV datasets. Best results are shown in boldface.}}\label{tab-RG}
		{\begin{tabular}{|c|l|cccc|}
			\hline
			Dataset&Variants&PSNR&SSIM&SNR&CR\\\hline
			\multirow{3}*{SAITD}&DLCM&26.37 &	0.725&0.664&14.200 \\
			&Conv&27.92 &	0.798 &\textbf{0.678}&14.250 \\
			&CD-Conv&\textbf{28.17}& \textbf{0.807} &\textbf{0.678}&\textbf{14.259} \\
			\hline
			\multirow{3}*{Hui}&DLCM&32.32 &	0.832 &0.820&15.167 \\
			&Conv&33.00 &	0.854 &	0.846&15.198 \\
			&CD-Conv&\textbf{33.12} &	\textbf{0.857} &\textbf{0.859}&\textbf{15.203 }\\
			\hline
			\multirow{3}*{Anti-UAV}&DLCM&31.44 &	0.901 &0.946&6.739 \\
			&Conv&\textbf{31.85} &	0.913 &0.960&6.696 \\
			&CD-Conv&\textbf{31.85} &	\textbf{0.914} &\textbf{0.965}&\textbf{6.709} \\
			\hline
			\multirow{3}*{Avg.}&DLCM&30.04 &	0.820&0.810&12.035 \\
			&Conv&30.93 &	0.855 &	0.828&12.048 \\
			&CD-Conv&\textbf{31.05}&	\textbf{0.859} &\textbf{0.834}&\textbf{12.057} \\
			\hline
		\end{tabular}}
	\end{table}
	where $\sigma_b$ is the standard deviation of the background area. SCRG is used to measure the SCR improvement of detection algorithms and is formulated as:
	\begin{equation}\label{eq13}
		\begin{split}
			f_{\rm{SCRG}} &= \frac{{\rm{SCG}}^{out}}{{\rm{SCG}}^{in}}= \frac{|\mu_t^{out}-\mu_b^{out}|\ / \sigma_b^{out}}{|\mu_t^{in}-\mu_b^{in}|\ / \sigma_b^{in}},
		\end{split}
	\end{equation}
	where $\mu_t$ and $\mu_b$ are the mean value of the target area and the background area respectively. CG is used to measure the improvement of contrast between targets and background and is formulated as:
	\begin{align}\label{eq14}
		f_{\rm{CG}} = \frac{{\rm{CR}}_{out}}{{\rm{CR}}_{in}} = \frac{|\mu_t^{out}-\mu_b^{out}|}{|\mu_t^{in}-\mu_b^{in}|}.
	\end{align}
	{Note that, in order to avoid the value of ``Inf" (\textit{i.e.,} the denominator is zero) and ``NAN'' (\textit{i.e.,} the numerator and denominator are both zero), we add $\epsilon$ to each denominator in equations \ref{eq11}-\ref{eq14} to prevent it from being zero. $\epsilon$ is set to $1e-10$ in our paper. ROC is used to measure the trend between detection probability $P_d$ and false alarm probability $F_a$, which are formulated as:
	\begin{align}
		P_d=\frac{{{\rm{TD}}}}{{{\rm{AT}}}},\\
		F_a=\frac{{{\rm{FD}}}}{{{\rm{NP}}}},
	\end{align}
	where $\rm{TD}$ and $\rm{FD}$ are the number of true detection and false detection. $\rm{AT}$ and $\rm{NP}$ are the amount of targets and the number of image pixels. Note that, the criterion for judging true detection is that the distance between the detected location and the groundtruth location is less than threshold $\tau$ and $\tau$ is set to 10 pixels \cite{SAITD} in our paper.}

	\subsubsection{Network Parameters}\label{sec-net_params}
	{The parameters of CD-RG in the feature extraction is CD-RG($D$=4, $K$=6, $G$=32) and the parameters of RGs are RG1,2($D$=1, $K$=4, $G$=64), RG3($D$=8, $K$=6, $G$=32). The parameters of the two LSTAs are LSTA1($kern$=3, $dila$=3) and LSTA2($kern$=3, $dila$=1).}

	\subsubsection{Training Details}\label{sec-training}
	{
		During the training phase, we randomly extracted 7 consecutive frames from an LR video clip, and randomly cropped a 64$\times$64 patch as the input. Meanwhile, its corresponding patch in HR video clip was cropped as groundtruth. We followed \cite{SOFVSR20} to augment the training data by random flipping and rotation.}

	{All experiments were implemented on a PC with an Nvidia RTX 3090 GPU. The networks were optimized using the Adam method \cite{Adam} with $\lambda_1$ = 0.9, $\lambda_2$ = 0.999 and the batch size was set to 12. The learning rate was initially set to $1e-3$ and halved in 10K, 20K, 60K iterations. We trained our network from scratch for 100K iterations.}

\begin{figure}[t]
	\centering
	\includegraphics[width=9cm]{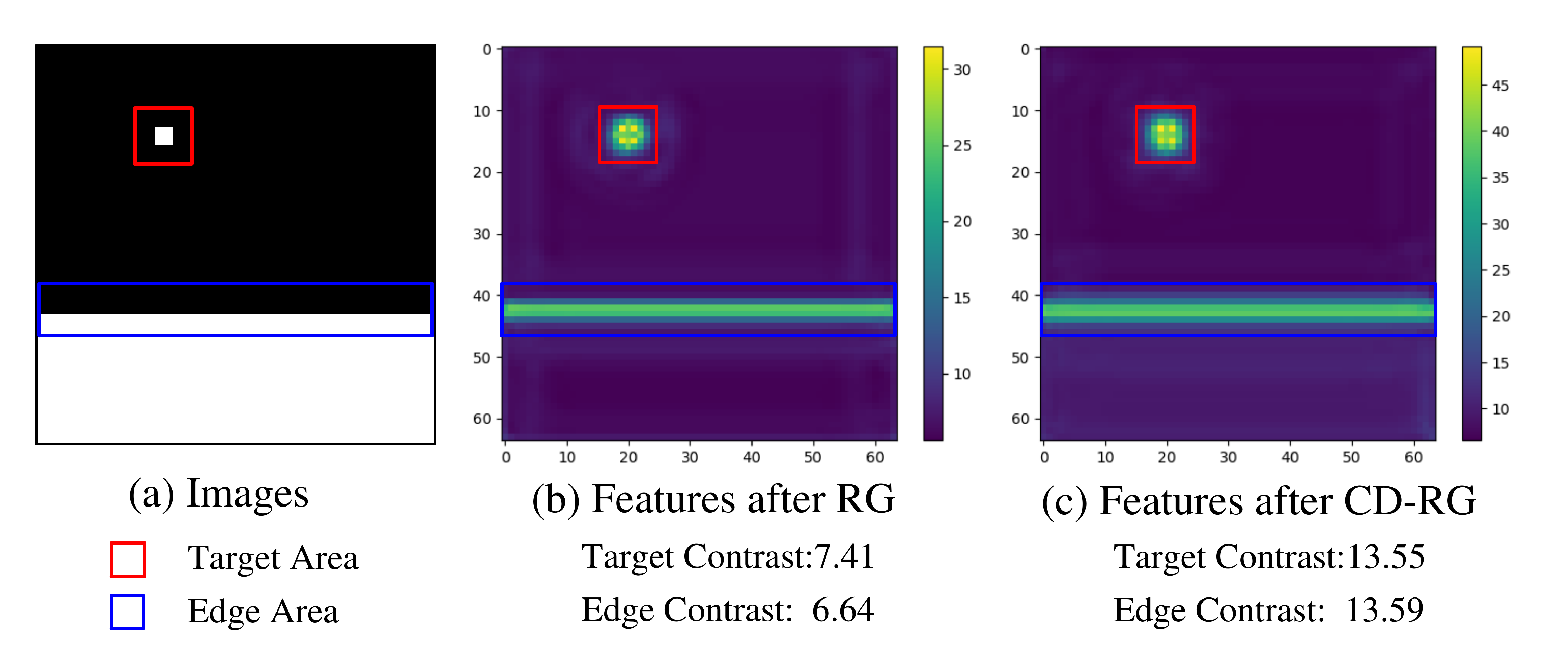}
	\caption{{A toy example of features generated by RG (b) and CD-RG (c). Note that, (a) represents the corresponding frame of the input image sequence. Red and blue boxes represent target and edge area, and the remaining area is background area.}}\label{cd_RG}
\end{figure}

	\subsection{Ablation Study}\label{ablation}
	In this subsection, we conduct ablation experiments to validate our design choice.

	\subsubsection{{Central Difference Residual Group}}\label{CD-RG_abl}
	{To demonstrate the effectiveness of our central difference residual group (CD-RG), we replace all the CD-Convs in CD-RG by Convs (\textit{i.e.,} residual group) and retrain the network from scratch. The experimental results in Table~\ref{tab-RG} show that CD-RG (\textit{i.e.,} CD-Convs) can introduce 0.12dB/0.004 gains on PSNR/SSIM and 0.06/0.09 gains on SNR/CR. This demonstrates that CD-RG can exploit the spatial local contrast prior to effectively improve the SR performance and the target contrast.}

	{In addition, we visualize the feature maps generate by residual group (RG) and CD-RG with a toy example in Fig.~\ref{cd_RG}. Note that, the visualization maps are the L2 norm results along the channel dimension \cite{visual111,DNAnet} and the red and blue boxes represent target areas and edge areas respectively. As is illustrated in Fig.~\ref{cd_RG}(a), the input frame of the image sequence consists of a target of size 3$\times$3 (\textit{i.e.,} the white cube at the top) and the clutter (\textit{i.e.,} the white area at the bottom). It can be observed from Figs.~\ref{cd_RG}(b) and (c) that the target contrast in the feature map extracted by CD-RG is higher than that of RG. This demonstrates that CD-RG can enhance the target contrast (from 7.41 to 13.55). In addition, CD-RG can also improve the contrast between high-frequency edges and background (from 6.64 to 13.59). This is because, CD-RG aggregates the gradient-level information to concentrate more on the high-frequency edge information, thus improving the SR performance and target contrast simultaneously.}
	%the values of the high-intensity clutter and the low-intensity background in the feature map extracted by CD-RG are similar while the values of those in the feature map extracted by RG are different. This is because, CD-RG aggregates the gradient-level detailed message to suppress background clutter while RG aggregates intensity-level semantic information.}

	{Moreover, we conduct ablation experiments to replace all the CD-convs in MoCoPnet by DLCMs. Note that, the training process of MoCoPnet with DLCMs is unstable with sudden loss divergence due to gradient fracture. By contrast, CD-conv reserves the image feature information to update all pixels, which ensures the gradient propagation continuity. The ablation results in Table~\ref{tab-RG} show that CD-conv introduces significant performance gain on PSNR/SSIM (\textit{i.e.,} 1.01/0.039 on average) and further improve the contrast of small targets (\textit{i.e.,} 0.024/0.022 SNR/CR gain on average).}

%	\begin{table}
%	\scriptsize
%	\renewcommand\arraystretch{1.2}
%	\centering
%	\caption{\scriptsize{{Ablation results of the local spatio-temporal attention module on the average of SAITD, Hui and Anti-UAV datasets. Note that, LSTA1 validates the effectiveness of the module and LSTA2-5 investigate the impact of its parameters, numbers, sub-pixel information exploitation and arrangements on SR performance. OFM, DAM validates the superiority of LSTA than optical flow technique and deformable alignment technique. ``C-LSTA[(a,b),(c,d)]'' represents the cascaded LSTA(a,b) and LSTA(c,d). ``P-LSTA[(a,b),(c,d)]'' represents the parallel LSTA(a,b) and LSTA(c,d). Best results are shown in boldface.}}}\label{tab-LSTA}
%{\begin{tabular}{|l|l|p{0.5cm}<{\centering}p{0.5cm}<{\centering}p{0.4cm}<{\centering}p{0.7cm}<{\centering}|}
%	\hline
%	Variants&Details&PSNR&SSIM&SNR&CR\\\hline
%	LSTA$_1$&\scriptsize{w/o LSTA}&30.77 &	0.851 &0.813&12.044 \\
%	LSTA$_2$&\scriptsize{C-LSTA[(3,1),(3,1)]}&31.02 &0.857 &0.831&12.054 \\
%	LSTA$_3$&\scriptsize{LSTA(3,1)}&30.89 &0.854 &0.825&12.051 \\
%	LSTA$_4$&\scriptsize{LSTA(9,1/4)}&30.96 &0.855 &0.828&12.051 \\
%	LSTA$_5$&\scriptsize{P-LSTA[(3,1),(3,3),(3,5)]}&30.98 &	0.857 &0.829&12.052 \\
%	OFM&\scriptsize{Optical Flow Module}&30.94&0.855&0.819&12.048 \\
%	DAM&\scriptsize{Deformable Alignment Module}&30.99 &	0.857 &	0.829 &	12.051 \\
%	MoCoPnet&\scriptsize{C-LSTA[(3,3),(3,1)]}&\textbf{31.05} &	\textbf{0.859} &\textbf{0.834}&\textbf{12.057} \\
%	\hline
%\end{tabular}}
%\end{table}

	\begin{table}
	\scriptsize
	\renewcommand\arraystretch{1.2}
	\centering
	\caption{\scriptsize{{Ablation results of the local spatio-temporal attention module on the average of SAITD, Hui and Anti-UAV datasets. Note that, LSTA1 validates the effectiveness of the module and LSTA2-5 investigate the impact of its parameters, numbers, sub-pixel information exploitation and arrangements on SR performance. OFM, DAM validates the superiority of LSTA than optical flow technique and deformable alignment technique. ``\#Params.'' represents the number of parameters. FLOPs is computed based on HR frames with a resolution of 256$\times$256. Best results are shown in boldface.}}}\label{tab-LSTA}
	{\begin{tabular}{|l|c|c|c|c|c|c|}
			\hline
			Variants&\#Params.&FLOPs&PSNR&SSIM&SNR&CR\\\hline
			LSTA$_1$&\textbf{4.83M} &\textbf{52.60G}&30.77 &	0.851 &0.813&12.044 \\
			LSTA$_2$&\textbf{4.83M} &52.65G&31.02 &0.857 &0.831&12.054 \\
			LSTA$_3$&\textbf{4.83M} &52.62G&30.89 &0.854 &0.825&12.051 \\
			LSTA$_4$&\textbf{4.83M} &52.62G&30.96 &0.855 &0.828&12.051 \\
			LSTA$_5$&4.84M &52.98G&30.98 &	0.857 &0.829&12.052 \\
			OFM&4.94M &55.35G&30.94&0.855&0.819&12.048 \\
			DAM&5.02M&56.45G&30.99 &0.857 &	0.829 &	12.051 \\
			MoCoPnet&\textbf{4.83M }&52.65G&\textbf{31.05} &	\textbf{0.859} &\textbf{0.834}&\textbf{12.057} \\
			\hline
	\end{tabular}}
\end{table}

\begin{figure}[t]
	\centering
	\includegraphics[width=9cm]{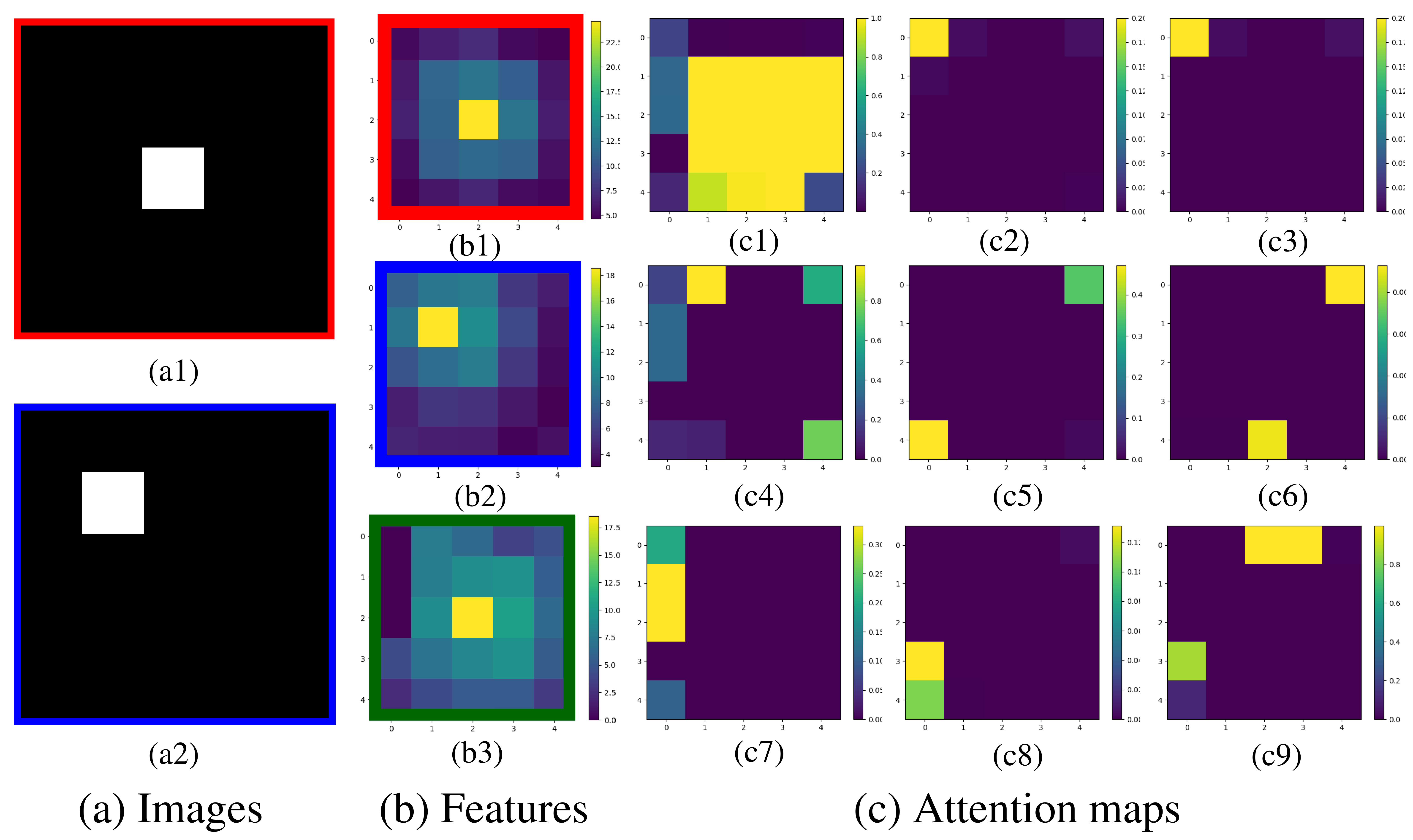}
	\caption{{A toy example illustration of feature maps and attention maps generated by LSTA$_3$. (a1), (a2) represent reference frame and neighborhood frame.  (b1), (b2) represent the reference feature and neighborhood feature before LSTA. (b3) represent the aligned feature after LSTA. (c1)-(c9) represent the attention maps of LSTA and the  positions correspond to the spatial arrangement in Fig.~\ref{fig-subpixel}(b).}}\label{fig_att_vsi}
\end{figure}

	{\subsubsection{Local Spatio-Temporal Attention Module}\label{sec-ablaLSTA}
	In MoCoPnet, two cascaded LSTAs with parameters LSTA($kern$=3, $dila$=3) and LSTA($kern$=3, $dila$=1) are used to enhance the spatio-temporal local features of sequence images in a coarse-to-fine manner. To validate the effectiveness of our design choice, we first remove LSTAs in MoCoPnet and name the model as LSTA$_1$. In addition, we further conduct ablation experiments to investigate the influences of the parameters, numbers, sub-pixel information exploitation and arrangements of LSTAs on SR performance. Specifically, we first replace LSTAs in MoCoPnet by two cascaded LSTAs with parameters ($kern$=3, $dila$=1) and name the model as LSTA$_2$. Secondly, we replace LSTAs in MoCoPnet by an LSTA with parameter ($kern$=3, $dila$=1) and name the model as LSTA$_3$. {Thirdly, we replace LSTAs in MoCoPnet by an LSTA with parameter ($kern$=9, $dila$=1/4) and name the model as LSTA$_4$.} Fourthly, we replace LSTAs in MoCoPnet by three parallel LSTAs with parameters ($kern$=3, $dila$=1), ($kern$=3, $dila$=3), ($kern$=3, $dila$=5) and name the model as LSTA$_5$.}

\begin{figure}[t]
	\centering
	\includegraphics[width=9cm]{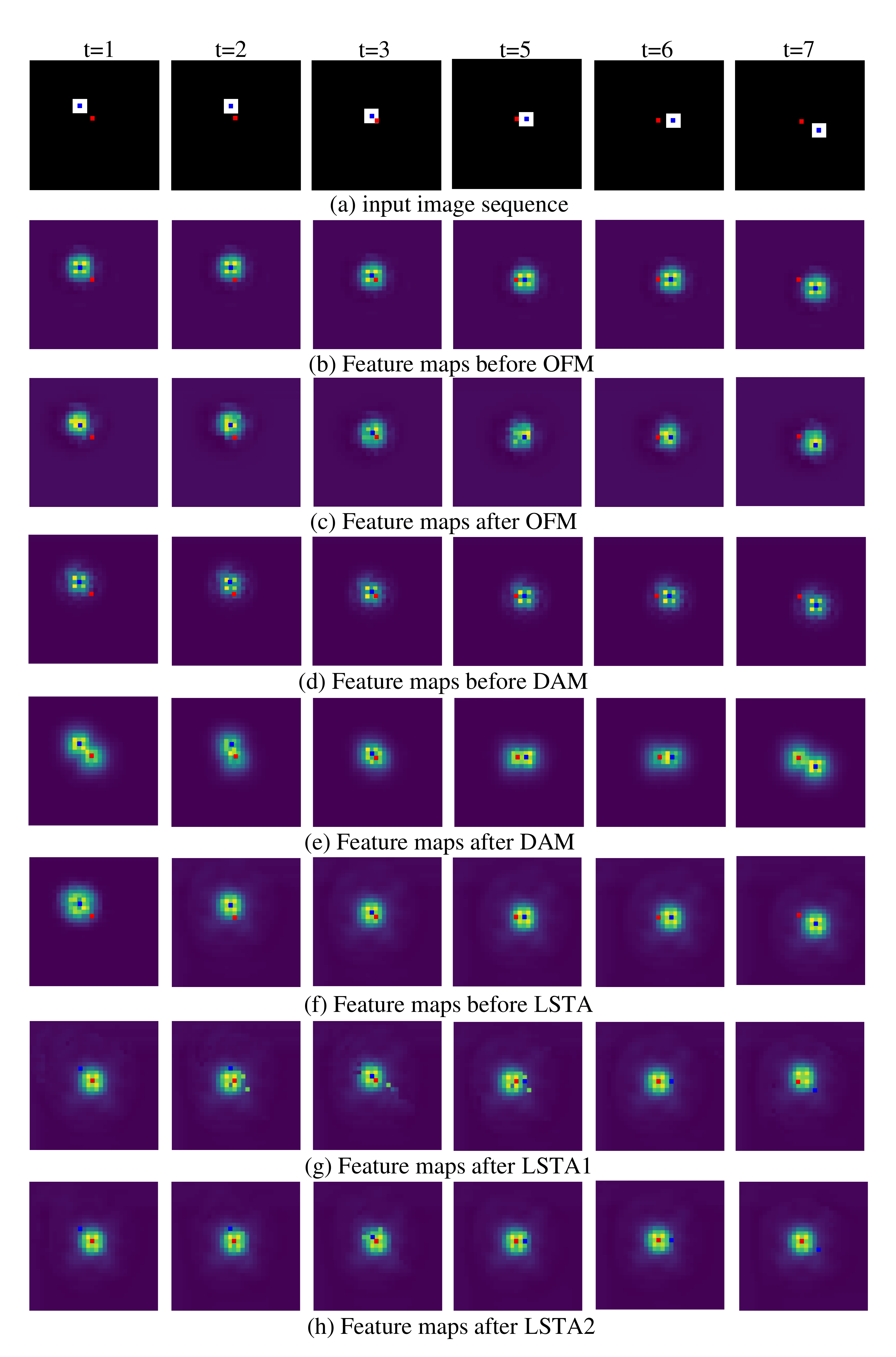}
	\caption{{A toy example illustration of feature maps generated by OFM, DAM and LSTA. Note that, $t$ represents the temporal dimension. The blue dot and the red dot represent the groundtruth position of the target in the current feature ($t\in[1,3]\cup[5,7]$) and in the reference feature ($t=4$).}}\label{fig_optical_v}
\end{figure}

\begin{table*}
	\scriptsize
	\centering
	\renewcommand\arraystretch{1.3}
	\caption{{\scriptsize{PSNR/SSIM results and running time of different methods achieved by SAITD \cite{SAITD}, Hui \cite{infrareddata} and Anti-UAV \cite{Anti-UAV} datasets. Note that, the running time is the total time tested on 100 consecutive frames with input resolution of 64$\times$64. Best results are shown in boldface.}}}\label{quantitative1}
	\scriptsize
	{\begin{tabular}{|l|c|c|c|c|c|c|c|c|c|c|}
		\hline
		{Methods}& {VSRnet} \cite{VSRnet}&{VESPCN} \cite{VESPCN}&{RCAN} \cite{RCAN}&{SOF-VSR} \cite{SOFVSR20}&{TDAN} \cite{TDAN}&{D3Dnet}\cite{D3Dnet}&{IERN}\cite{IERN}&{PSRGAN}\cite{PSRGAN}&{ChaSNet}\cite{ChaSNet}&{MoCoPnet}\\
		\hline
		SAITD&	26.03/0.706 &	26.57/0.735 &	26.58/0.735 &	26.97/0.753 &	26.11/0.709& 27.81/0.794 &26.53/0.730&26.49/0.722&26.71/0.760 &\textbf{28.17/0.807} \\
		Hui &	32.03/0.828 &	32.33/0.835 &	32.44/0.836 &	32.55/0.841 &	32.17/0.830 &	32.84/0.850 &32.67/0.838&32.36/0.828&32.41/0.824 &
			\textbf{33.12/0.857} \\
		Anti-UAV &	31.42/0.904&31.63/0.910 &	31.73/0.912 &	31.68/0.912 &	31.58/0.905 &	31.81/0.911 &31.54/0.909&30.96/0.884&31.73/0.910 
		&	\textbf{31.85/0.914} \\
		\hline
		Average&	29.83/0.813 &	30.18/0.827 &	30.25/0.828&30.40/0.835 &	29.95/0.815 &	30.82/0.852 &30.25/0.826&29.94/0.811&30.28/0.831 
		&	\textbf{31.05/0.859} \\\hline
		Time (s) &\textbf{0.157}&0.627&11.174&3.788&5.923&7.042&6.578&0.823&13.073&10.343\\
		\hline
	\end{tabular}}
\end{table*}

\begin{table*}
	%		\footnotesize
	\centering
	\renewcommand\arraystretch{1.3}
	\centering
	\caption{{SNR and CR results of different methods achieved on super-resolved LR images (columns $2^{nd}-9^{th}$) and super-resolved HR images (columns $10^{th}-17^{th}$). Note that, we add the results of LR and HR as the baseline results and the resolution of LR is 4 times lower than the listed resolution. Exclude LR and HR, best results are shown in boldface and second best results are shown in underlined.}}\label{tab-enhance}
	\scriptsize
{\begin{tabular}{|l|cc|cc|cc|cc|cc|cc|cc|cc|}
		\hline
		Resolution&\multicolumn{2}{c|}{640$\times$512}&\multicolumn{2}{c|}{256$\times$256}&\multicolumn{2}{c|}{640$\times$640}&\multicolumn{2}{c|}{-}&\multicolumn{2}{c|}{2560$\times$2048}&\multicolumn{2}{c|}{2048$\times$2048}&\multicolumn{2}{c|}{2560$\times$2048}&\multicolumn{2}{c|}{-}\\\hline
		\multirow{2}*{Methods}&\multicolumn{2}{c|}{SAITD}&\multicolumn{2}{c|}{Hui}&\multicolumn{2}{c|}{Anti-UAV}&\multicolumn{2}{c|}{Average}&\multicolumn{2}{c|}{SAITD}&\multicolumn{2}{c|}{Hui}&\multicolumn{2}{c|}{Anti-UAV}&\multicolumn{2}{c|}{Average}\\\cline{2-17}
		&SNR&CR&SNR&CR&SNR&CR&SNR&CR&SNR&CR&SNR&CR&SNR&CR&SNR&CR\\\hline
		LR&0.666&14.066&0.781&13.583&0.915&6.174&0.787&11.274&-&-&-&-&-&-&-&- \\
		Bicubic&\underline{0.676}&13.780&0.750&15.100&0.817&\textbf{6.747}&0.747&11.875&0.808&\textbf{14.736}&0.986&\textbf{15.817}&0.958&7.173&0.917&\textbf{12.575} \\
		VSRnet \cite{VSRnet}&0.659&14.125&0.776&15.100&0.882&6.726&0.773&11.984&0.672&14.704&1.002&15.661&0.954&\underline{7.178}&0.876&12.514 \\
		VESPCN \cite{VESPCN}&0.656&14.118&0.793&15.145&0.920&6.641&0.790&11.968&0.901&14.664&\underline{1.005}&15.616&0.963&7.115&0.956&12.465 \\
		RCAN \cite{RCAN}&0.670&14.213&0.813&15.202&0.952&6.713&0.811&12.043&\underline{0.914}&14.720&0.997&15.649&0.947&7.168&0.953&12.512 \\
		SOF-VSR \cite{SOFVSR20}&0.662&14.175&0.808&15.113&0.932&6.698&0.800&11.995&0.913&14.700&0.997&15.603&\underline{0.965}&7.168&0.958&12.490 \\
		TDAN \cite{TDAN}&0.655&14.206&0.772&15.192&0.882&6.711&0.770&12.036&0.889&14.725&0.999&15.693&0.963&7.173&0.950&12.530 \\
		D3Dnet \cite{D3Dnet}&0.672&\underline{14.240}&{0.845}&\textbf{15.215}&\textbf{0.972}&\underline{6.736}&\underline{0.830}&\textbf{12.064}&0.907&\underline{14.731}&\underline{1.005}&\underline{15.699}&0.964&7.166&0.959&\underline{12.532} \\
		{IERN}\cite{IERN}&0.665&13.582&0.833&15.155&0.888&6.537&0.795&11.758&0.887&14.320&0.934&15.683&0.960&7.162&0.927&12.389 
		\\
		{PSRGAN}\cite{PSRGAN}&\textbf{0.678}&13.767&\underline{0.852}&14.939&0.941&6.326&0.824&11.677&0.879&14.637&0.942&15.604&0.955&7.158&0.925&12.466 
		\\
		{ChaSNet}\cite{ChaSNet}&0.668&13.491&0.793&14.605&0.885&6.050&0.782&11.382&0.899&14.616&0.937&15.623&0.959&7.161&0.932&12.466 
		\\
		MoCoPnet&\textbf{0.678}&\textbf{14.259}&\textbf{0.859}&\underline{15.203}&\underline{0.965}&6.709&\textbf{0.834}&\underline{12.057}&\textbf{0.922}&14.729&\textbf{1.006}&15.685&\textbf{0.966}&\textbf{7.181}&\textbf{0.965}&\underline{12.532} \\
		HR&0.810&14.262&1.001&15.265&0.959&6.706&0.923&12.078&-&-&-&-&-&-&-&- \\
		\hline
	\end{tabular}}
\end{table*}

\begin{figure*}[t]
	\centering
	\includegraphics[width=18cm]{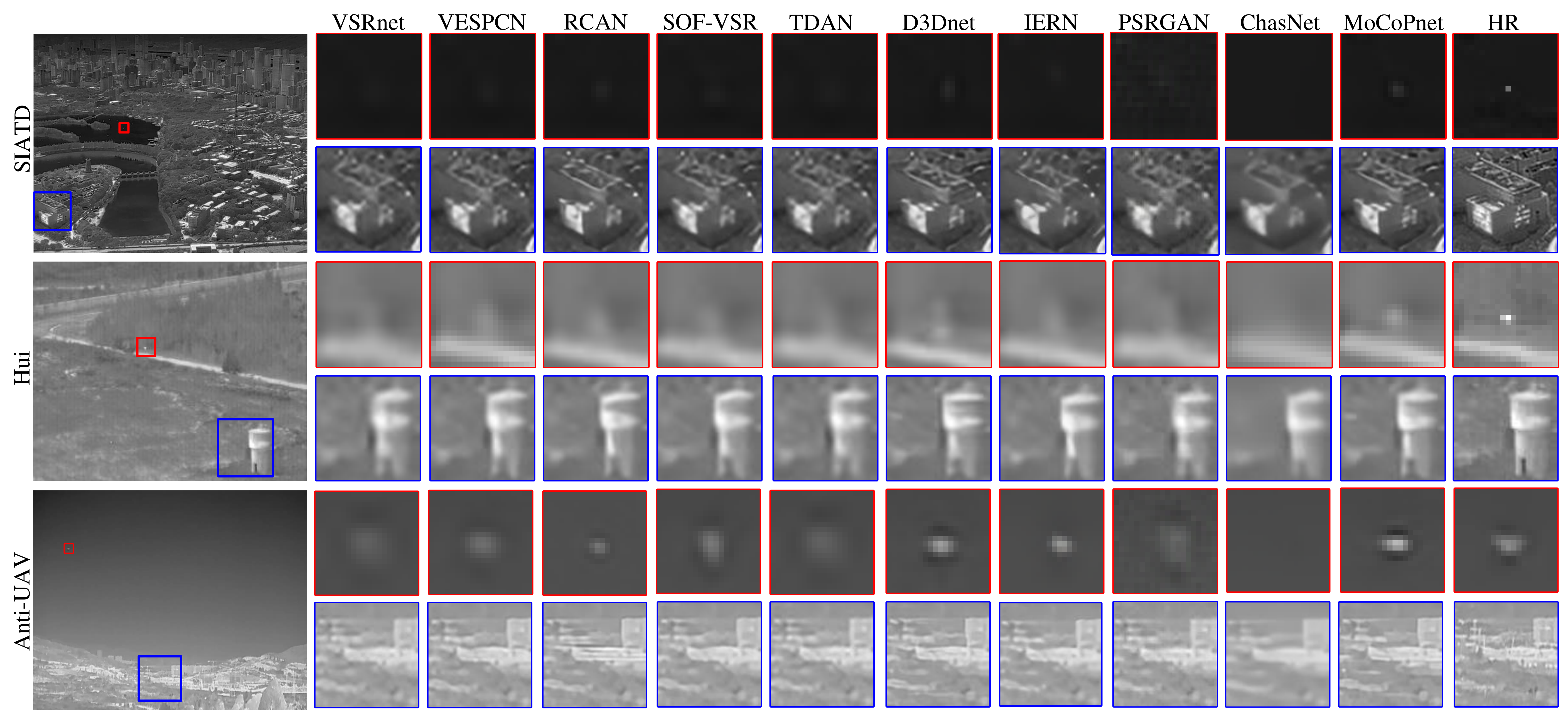}
	\caption{{Visual results of different SR methods on LR images for 4$\times$ SR.}}\label{fig_enhance}
\end{figure*}

\begin{figure*}[t]
	\centering
	\includegraphics[width=18cm]{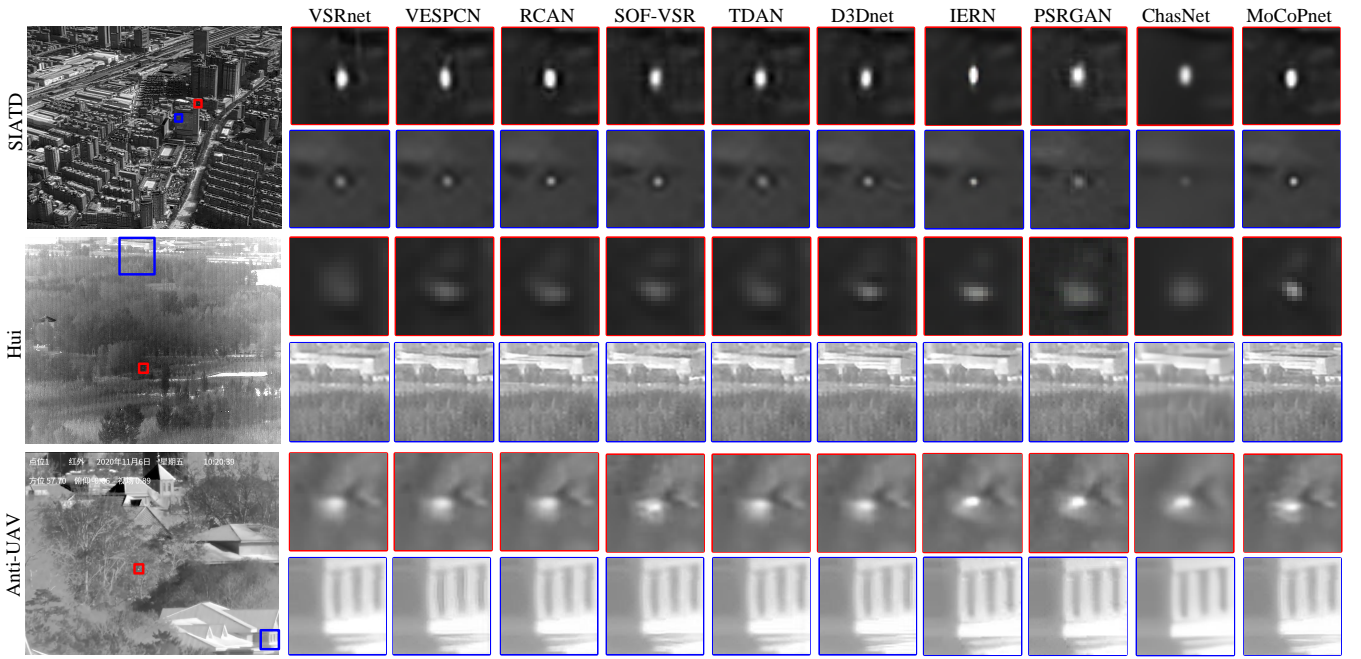}
	\caption{{Visual results of realSR on HR images for 4$\times$ SR.}}\label{fig_realSR}
	\vspace{.01pt}
\end{figure*}

	{The experimental results of LSTA$_{1-5}$ are shown in Table~\ref{tab-LSTA}. It can be observed that the \#Params. and FLOPs of LSTA1-5 are comparable, which means that LSTA only costs a small amount of computations. The PSNR/SSIM/SNR/CR scores of LSTA$_{1}$ are 0.28dB/0.008/0.021/0.013 lower than MoCoPnet. This demonstrates that LSTA can effectively use the supplementary temporal information to enhance the local features, thus improving the SR performance and the target contrast. The PSNR/SSIM/SNR/CR scores of LSTA$_{2}$ are 0.03dB/0.002/0.003/0.003 lower than MoCoPnet. This demonstrates that LSTA with larger expansion rate (\textit{i.e.,} $dila$=3) for coarse processing promotes our network to better extract and utilize temporal information. The PSNR/SSIM/SNR/CR scores of LSTA$_{3}$ are 0.16dB/0.004/0.009/0.006 lower than MoCoPnet and 0.13dB/0.003/0.006/0.003 lower than LSTA$_{2}$. This demonstrates that coarse-to-fine processing benefits SR performance and target enhancement. The PSNR/SSIM/SNR/CR scores of LSTA$_{4}$ are slightly higher than LSTA$_{3}$ for 0.07dB/0.001/0.003/0.000 but the memory cost of LSTA$_{4}$ is 2 times than LSTA$_{3}$ (\textit{i.e.,} 2.46 vs. 1.17). This demonstrates sub-pixel information exploitation benefits the performance of SR and target enhancement but significantly increases the memory cost. The PSNR/SSIM/SNR/CR scores of LSTA$_{5}$ are 0.07dB/0.002/0.005/0.005 lower than MoCoPnet and 0.09dB/0.003/0.004/0.001 higher than LSTA$_{3}$. This demonstrates that the cascade mode of LSTAs can better exploit inter-frame information correlation and SR performance and target enhancement can be further improved by enlarging the receptive field of LSTAs.}

\begin{table}
	\begin{center}
		\centering
		\renewcommand\arraystretch{1.4}
		\caption{\scriptsize{Parameter settings of Top-hat \cite{tophat}, ILCM \cite{ILCM} and IPI \cite{IPI} in HR images. ``{B}'' represents block size and ``{S}'' represents stride.}}\label{excel_detection}
		\scriptsize \vspace{-.17in}
		{\begin{tabular}{p{9.3mm}<{\centering}p{73.3mm}<{\centering}}
			\toprule[1pt]
			Method&Parameters\\
			\midrule
			Top-hat&5$\times$5 square filter\\
			\rule{0pt}{12pt}
			ILCM&5$\times$5 filter\\
			\rule{0pt}{11pt}
			IPI&B=50$\times$50, S=10, $\lambda=\frac{L}{\sqrt{\min(n_1,n_2,n_3)}}$, L=1, $\varepsilon$=$10^{-7}$\\
			\bottomrule[1pt]
		\end{tabular}}
	\end{center}
\end{table}

	{Note that, we visualize the feature maps and attention maps generated by LSTA$_3$ (\textit{i.e.,} an LSTA with kernel size of 3 and dilation rate of 1) with a toy example in Fig.~\ref{fig_att_vsi}. Note that, the visualized feature maps are the L2 norm results along the channel dimension \cite{visual111,DNAnet}. As is illustrated in Fig.~\ref{fig_att_vsi}(a1), the target with size 1$\times$1 (\textit{i.e.,} the white cube) is in the middle of the red reference frame. In Fig.~\ref{fig_att_vsi}(a2), the target is in the top left of the blue neighborhood frame. The corresponding features before LSTA are shown in Figs.~\ref{fig_att_vsi}(b1) and (b2). The aligned feature after LSTA is shown in Fig.~\ref{fig_att_vsi}(b3). It can be observed that LSTA can effectively perform frame alignment to achieve motion compensation. In addition, the attention maps are shown in Figs.~\ref{fig_att_vsi}(c1)-(c9), and the position of each attention map corresponds to the spatial arrangement in Fig.~\ref{fig-subpixel}(b). It can be observed that Fig.~\ref{fig_att_vsi}(c1) has the highest intensity (more than 90\% are 1) and represents the top-left motion, which demonstrates that LSTA can effectively capture the target motion to perform frame alignment.}

	{Finally, we replace LSTAs in MoCoPnet by an optical-flow module (OFM) and a deformable alignment module (DAM) to compare our LSTA with the widely used optical flow and deformable alignment techniques. The experimental results are listed in Table~\ref{tab-LSTA}. It can be observed that the PSNR/SSIM/SNR/CR scores of MoCoPnet with LSTAs are higher than MoCoPnet with OFM and DAM for 0.11dB/0.004/0.015/0.009 and 0.06dB/0.002/0.005/0.006 respectively. By contrast, the number of parameters and FLOPs of MoCoPnet with LSTA modules are lower than MoCoPnet with OFM and DAM for  0.11M/2.70G and 0.19M/3.80G respectively. This demonstrates that LSTA is superior in exploiting the information among frames to improve the SR performance and the target contrast with lower computational cost. This is because, on the one hand, LSTA can direct learn motion compensation by attention mechanism without optical flow estimation and warping, which results in ambiguous and duplicate results \cite{vague1,vague2}. On the other hand, compared with DAM, LSTA can better incorporate the local prior to achieve improved SR performance and the training process of LSTA is more stable to converge to a good results.}

	{In addition, we visualize the feature maps generated by OFM, DAM and LSTAs with a toy example in Fig.~\ref{fig_optical_v}. Note that, the visualization maps are the L2 norm results along the channel dimension. As is illustrated in Fig.~\ref{fig_optical_v}(a), the input image sequence consists of a random consistent movement of a target with size 3$\times$3 (\textit{i.e.,} the white cube) in the background (\textit{i.e.,} the black area). The feature maps before OFM, DAM and LSTAs are shown in Figs.~\ref{fig_optical_v}(b), (d) and (f). It can be observed that the target positions in the extracted feature maps are close to the blue dots (\textit{i.e.,} the groundtruth position of the target in the current feature). Then OFM, DAM and LSTAs perform feature alignment on the extracted features. As is illustrated in Fig.~\ref{fig_optical_v}(c), the target positions in the feature maps generated by OFM are close to the blue dots. In Fig.~\ref{fig_optical_v}(e), the blue dots and the red dots (\textit{i.e.,} the groundtruth position of the target in the reference feature) are both highlighted, which demonstrates that DAM does not perform frame alignment but highlight all the possible positions. The feature maps generated by LSTA1($kern$=3, $dila$=3) and LSTA2( $kern$=3, $dila$=1) are shown in Figs.~\ref{fig_optical_v}(e) and (f). As is illustrated in Fig.~\ref{fig_optical_v}(f), all the target positions in the feature maps generated by LSTA2 are closer to the red dot than those of OFM. This demonstrates that LSTA is superior in motion compensation. Note that, it can be observed from Figs.~\ref{fig_optical_v}(e) and (f) that LSTA1 and LSTA2 achieve coarse-to-fine alignment to highlight the aligned target. This demonstrates the effectiveness and superiority of our coarse-to-fine alignment strategy.}

	{\subsection{Comparative Evaluation}
	{In this subsection, we compare our MoCoPnet with 1 top-performing single image SR methods {RCAN} \cite{RCAN}, 5 video SR methods {VSRnet} \cite{VSRnet}, {VESPCN} \cite{VESPCN}, {SOF-VSR} \cite{SOFVSR20} and {TDAN} \cite{TDAN}, D3Dnet\cite{D3Dnet} and 3 infrared image SR methods {IERN} \cite{IERN}, {PSRGAN} \cite{PSRGAN}, and ChaSNet \cite{ChaSNet}. For fair comparison, we retrain all the compared methods on infrared small target dataset \cite{infrareddata} and exclude the first and the last 2 frames of the video sequences for performance evaluation. }}

\begin{table*}
	\centering
	\renewcommand\arraystretch{1.3}
	\caption{\scriptsize{Quantitative detection results of Tophat, ILCM and IPI achieved on super-resolved LR images in infrared small target datasets. Best results are shown in boldface and second best results are shown in underlined.} %``Inf" represents infinity and ``NAN'' represents the numerator and denominator are both zero.
	}\label{excel_dectectLR}
	\scriptsize
%	\vspace{7pt}
	{\begin{tabular}{|c|l|p{5.4mm}<{\centering}p{5.4mm}<{\centering}p{5.4mm}<{\centering}p{5.4mm}<{\centering}|p{5.4mm}<{\centering}p{5.4mm}<{\centering}p{5.4mm}<{\centering}p{5.4mm}<{\centering}|p{5.4mm}<{\centering}p{5.4mm}<{\centering}p{5.4mm}<{\centering}p{5.4mm}<{\centering}|p{5.4mm}<{\centering}p{5.4mm}<{\centering}p{5.4mm}<{\centering}p{5.4mm}<{\centering}|}
		\hline
		\multicolumn{2}{|c|}{Resolution}&\multicolumn{4}{c|}{640$\times$512}&\multicolumn{4}{c|}{256$\times$256}&\multicolumn{4}{c|}{640$\times$640}&\multicolumn{4}{c|}{-}\\\hline
		\multicolumn{2}{|c}{\multirow{2}*{Methods}}&\multicolumn{4}{|c|}{SAITD}&\multicolumn{4}{c|}{Hui}&\multicolumn{4}{c|}{Anti-UAV}&\multicolumn{4}{c|}{Average}\\\cline{3-18}
		\multicolumn{2}{|c|}{}&SNRG&BSF&SCRG&CG&SNRG&BSF&SCRG&CG&SNRG&BSF&SCRG&CG&SNRG&BSF&SCRG&CG\\
		\hline
		\multirow{4}*{Top-hat}&Bicubic&0.50&2.55&6.43&\underline{3.60}&0.93&1.79&15.94&9.66&3.77&4.32&15.58&3.12&1.73&2.89&12.65&5.46 \\
		&D3Dnet&0.77&\underline{3.10}&\textbf{9.31}&\textbf{4.28}&1.49&\textbf{2.01}&\textbf{20.47}&\textbf{11.20}&\underline{9.60}&6.70&\underline{31.05}&\textbf{3.33}&\underline{3.95}&3.94&\textbf{20.28}&\textbf{6.27} \\
		&MoCoPnet&\underline{0.82}&\textbf{3.25}&\underline{9.04}&3.55&\underline{1.53}&\underline{1.98}&\underline{18.88}&\underline{10.22}&\textbf{13.06}&\underline{6.72}&28.35&2.82&\textbf{5.14}&\underline{{3.99}}&\underline{{18.76}}&\underline{{5.53}} \\
		&HR&\textbf{1.62}&1.84&5.40&3.22&\textbf{1.73}&1.55&8.82&5.30&7.61&\textbf{13.18}&\textbf{74.73}&\underline{2.99}&3.65&\textbf{5.52}&29.65&3.83 \\
		\hline
		\multirow{4}*{ILCM}&Bicubic&1.07&\textbf{1.20}&5.93&4.89&0.96&\textbf{0.91}&3.73&4.34&0.90&\textbf{0.83}&1.82&2.15&0.97&\textbf{0.98}&3.83&3.79 \\
		&D3Dnet&1.07&\underline{1.02}&{7.29}&7.15&1.08&\underline{0.84}&8.01&10.03&\textbf{1.07}&\underline{0.77}&2.84&3.72&1.07&\underline{0.87}&6.05&6.96 \\
		&MoCoPnet&\underline{1.08}&1.00&\underline{7.91}&\underline{7.91}&\underline{1.09}&\underline{0.84}&\underline{8.21}&\underline{10.12}&\textbf{1.07}&0.76&\textbf{3.30}&\underline{4.47}&\underline{1.08}&\underline{0.87}&\underline{6.47}&\underline{7.50 }\\
		&HR&\textbf{1.37}&0.89&\textbf{11.85}&\textbf{13.20}&\textbf{1.31}&0.77&\textbf{10.88}&\textbf{15.39}&\underline{1.05}&0.70&\underline{3.14}&\textbf{4.54}&\textbf{1.24}&0.79&\textbf{8.62}&\textbf{11.04 }\\
		\hline
		\multirow{4}*{IPI}&Bicubic&5.1$e$7&\textbf{2.1$e$10}&1.6$e$7&1$e$-3&6.2$e$9&\textbf{1.1$e$10}&2.1$e$9&0.82 &9.8$e$9&\textbf{9.9$e$9}&3.3$e$9&0.06 &5.3$e$9&\textbf{1.4$e$10}&1.8$e$9&0.29 \\
		&D3Dnet&1.9$e$8&\underline{4.0$e$9}&2.1$e$8&0.07 &\underline{1.9$e$10}&\underline{5.0$e$9}&\underline{3.5$e$9}&\underline{0.96} &\underline{2.0$e$10}&3.2$e$9&\underline{4.1$e$9}&\underline{0.13 }&\underline{1.3$e$10}&4.1$e$9&\underline{2.6$e$9}&\underline{0.39} \\
		&MoCoPnet&\underline{3.5$e$8}&2.7$e$9&\underline{6.4$e$8}&\underline{0.11 }&\textbf{2.0$e$10}&4.5$e$9&\textbf{4.6$e$9}&\textbf{1.12 }&\textbf{4.2$e$10}&\underline{5.4$e$9}&\textbf{8.1$e$9}&\underline{0.13 }&\textbf{2.1$e$10}&\underline{4.2$e$9}&\textbf{4.4$e$9}&\textbf{0.45} \\
		&HR&\textbf{1.1$e$10}&2.6$e$9&\textbf{2.3$e$9}&\textbf{0.13} &0.60 &2.2$e$9&8.63 &0.61 &8.0$e$9&1.5$e$9&1.3$e$9&\textbf{0.16 }&6.2$e$9&2.1$e$9&1.2$e$9&0.30 \\\hline
	\end{tabular}}
\end{table*}

\begin{table*}
	\centering
	\renewcommand\arraystretch{1.3}
	\caption{\scriptsize{Quantitative detection results of Tophat, ILCM and IPI achieved on super-resolved HR images in infrared small target datasets. Best results are shown in boldface and second best results are shown in underlined. }}\label{excel_dectectHR}
	\scriptsize
	{\begin{tabular}{|c|l|p{5.4mm}<{\centering}p{5.4mm}<{\centering}p{5.4mm}<{\centering}p{5.4mm}<{\centering}|p{5.4mm}<{\centering}p{5.4mm}<{\centering}p{5.4mm}<{\centering}p{5.4mm}<{\centering}|p{5.4mm}<{\centering}p{5.4mm}<{\centering}p{5.4mm}<{\centering}p{5.4mm}<{\centering}|p{5.4mm}<{\centering}p{5.4mm}<{\centering}p{5.4mm}<{\centering}p{5.4mm}<{\centering}|}
		\hline
		\multicolumn{2}{|c|}{Resolution}&\multicolumn{4}{c|}{2560$\times$2048}&\multicolumn{4}{c|}{2048$\times$2048}&\multicolumn{4}{c|}{2560$\times$2560}&\multicolumn{4}{c|}{-}\\\hline
		\multicolumn{2}{|c}{\multirow{2}*{Methods}}&\multicolumn{4}{|c|}{SAITD}&\multicolumn{4}{c|}{Hui}&\multicolumn{4}{c|}{Anti-UAV}&\multicolumn{4}{c|}{Average}\\\cline{3-18}
		\multicolumn{2}{|c|}{}&SNRG&BSF&SCRG&CG&SNRG&BSF&SCRG&CG&SNRG&BSF&SCRG&CG&SNRG&BSF&SCRG&CG\\
		\hline
		\multirow{3}*{Top-hat}&Bicubic&1.01&\textbf{1.79}&4.45&2.78&1.19&\textbf{1.82}&\textbf{6.09}&\underline{3.03}&\textbf{4.66}&6.95&49.83&\textit{4.40}&\textbf{2.28}&\underline{3.52}&20.13&3.40\\
		&D3Dnet&\underline{1.27}&\underline{1.70}&\textbf{4.67}&\textbf{2.95}&\textbf{1.21}&1.74&5.63&2.98&\underline{3.21}&\textbf{7.39}&\textbf{59.37}&\underline{4.40}&1.90&\textbf{3.61}&\textbf{23.22}&\underline{3.44} \\
		&MoCoPnet&\textbf{1.33}&\underline{1.70}&\underline{4.60}&\underline{2.87}&\underline{1.20}&\underline{1.78}&\underline{5.88}&\textbf{3.11}&3.20&\underline{7.35}&\underline{58.17}&\textbf{4.53}&\underline{1.91}&\textbf{3.61}&\underline{22.88}&\textbf{3.50} \\
		\hline
		\multirow{3}*{ILCM}&Bicubic&1.20&\textbf{0.90}&9.02&11.10&\underline{0.97}&\textbf{0.84}&9.72&11.76&\underline{1.00}&\textbf{0.86}&7.77&8.71&1.06&\textbf{0.86}&8.84&10.52 \\
		&D3Dnet&\underline{1.37}&\underline{0.85}&\underline{12.65}&\underline{16.53}&\textbf{1.04}&\underline{0.78}&\textbf{10.54}&\underline{13.94}&\textbf{1.02}&\underline{0.85}&\underline{8.58}&\underline{9.60}&\underline{1.14}&\underline{0.82}&\underline{10.59}&\underline{13.35} \\
		&MoCoPnet&\textbf{1.40}&\underline{0.85}&\textbf{12.86}&\textbf{16.90}&\textbf{1.04}&0.76&\underline{10.50}&\textbf{14.18}&\textbf{1.02}&\underline{0.85}&\textbf{8.71}&\textbf{9.75}&\textbf{1.15}&\underline{0.82}&\textbf{10.69}&\textbf{13.61} \\
		\hline
		\multirow{3}*{IPI}&Bicubic&2.5$e$9&\textbf{3.9$e$9}&2.8$e$8&0.039 &\textbf{5.1$e$10}&\textbf{2.6$e$10}&1.3$e$10&0.14 &\textbf{1.7$e$9}&\textbf{1.2$e$9}&\textbf{3.0$e$8}&0.046 &\textbf{1.8$e$10}&\textbf{1.0$e$10}&4.6$e$9&0.074 \\
		&D3Dnet&\underline{3.4$e$9}&\underline{2.9$e$9}&\underline{3.1$e$8}&\textbf{0.048} &\underline{4.8$e$10}&\underline{1.9$e$10}&\underline{1.6$e$10}&\underline{0.16} &\underline{3.5$e$8}&\underline{4.0$e$7}&\underline{7.9$e$7}&\underline{0.048} &\underline{1.7$e$10}&\underline{7.2$e$9}&\underline{5.4$e$9}&\underline{0.086 }\\
		&MoCoPnet&\textbf{3.6$e$9}&2.6$e$9&\textbf{3.2$e$8}&\underline{0.044} &3.7$e$10&1.6$e$10&\textbf{1.7$e$10}&\textbf{0.18 }&0.23 &20.90 &9.95 &\textbf{0.049} &1.3$e$10&6.2$e$9&\textbf{5.9$e$9}&\textbf{0.092} \\
		\hline
	\end{tabular}}
\end{table*}

\begin{figure*}[t]
	\centering
	\includegraphics[width=18cm]{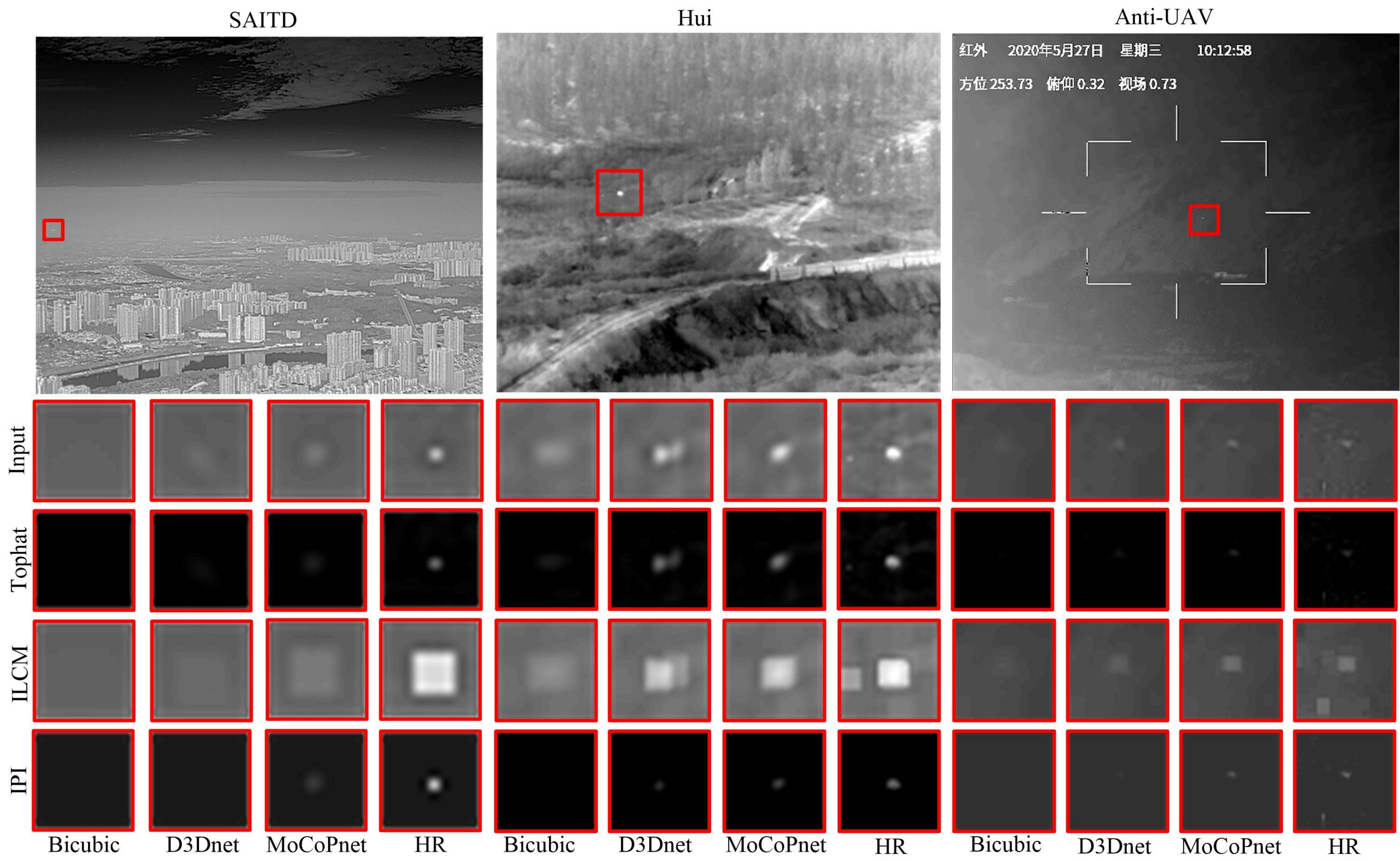}
	\caption{Qualitative results of super-resolved LR image and detection results in SAITD, Hui and Anti-UAV datasets.}\label{fig_18_256}%
\end{figure*}

	{\subsubsection{SR on Synthetic Images}
	{PSNR/SSIM results calculated on the whole image are listed in Table~\ref{quantitative1}. SNR and CR scores calculated in the local background neighborhood are listed in the $2^{nd}-9^{th}$ columns of Table \ref{tab-enhance}. It can be observed that MoCoPnet achieves the highest scores of PSNR, SSIM and outperforms most of the compared algorithms on SNR and CR scores. The above scores demonstrate that our network can effectively recover accurate details and improve the target contrast. That is because, LSTA performs implicit motion compensation and CD-RG incorporates the center-orient gradient information to effectively improve the SR performance and the target contrast.} {Note that, we also analyze the running time of different methods and the results are shown in Table~\ref{quantitative1}. The running time is the total time tested on 100 consecutive HR frames with a resolution of 256$\times$256 and is averaged over 20 runs. It can be observed that our MoCoPnet achieves better SR performance with a reasonable increase in running time. }}

	{Qualitative results are shown in Fig.~\ref{fig_enhance}. For SR performance, it can be observed from the blue zoom in regions that MoCoPnet can recover more accurate details  (\textit{e.g.,} the sharp edges of buildings, and the lighthouse details closer to groundtruth HR image). For target enhancement, it can be observed from the red zoom in regions that, in the first row, MoCoPnet can further improve the target contrast which is almost invisible in other compared methods. In the second row, MoCoPnet is more robust to large motion caused by turntable collections \cite{infrareddata} (\textit{e.g.,} artifacts in the zoom-in region of D3Dnet). In the third row, MoCoPnet can effectively improve the target contrast to be even higher than HR images (\textit{i.e.,} 1.82 vs. 1.75). }

	{\subsubsection{SR on Real Images}
	{SNR and CR scores calculated in the local background neighborhood of super-resolved HR images are listed in the $10^{th}-17^{th}$ columns of Table \ref{tab-enhance}. It can be observed that MoCoPnet can achieve the best SNR score and the second best CR score on the average of test datasets under real-world degradation. This demonstrates the superiority of our method in improving the contrast between targets and background. }}

	{Qualitative results are shown in Fig.~\ref{fig_realSR}. It can be observed that MoCoPnet can recover finer details and achieve better visual quality, such as the edges of building and window. In addition, MoCoPnet can further improve the intensity and the contour details of the targets.}  %In addition, the image reconstructed by MoCoPnet has less distortion and higher fidelity, which is more in line with the actual application requirements. Specifically, the zoom-in local neighborhood background of data20 in the D3Dnet super-resolved image has obvious noise, which can not meet the requirements of subsequent image processing such as detection and recognition.

	\subsection{Effect On Infrared Small Target Detection Algorithm}\label{Detection_effect}

	\begin{figure*}[t]
		\centering
		\includegraphics[width=18cm]{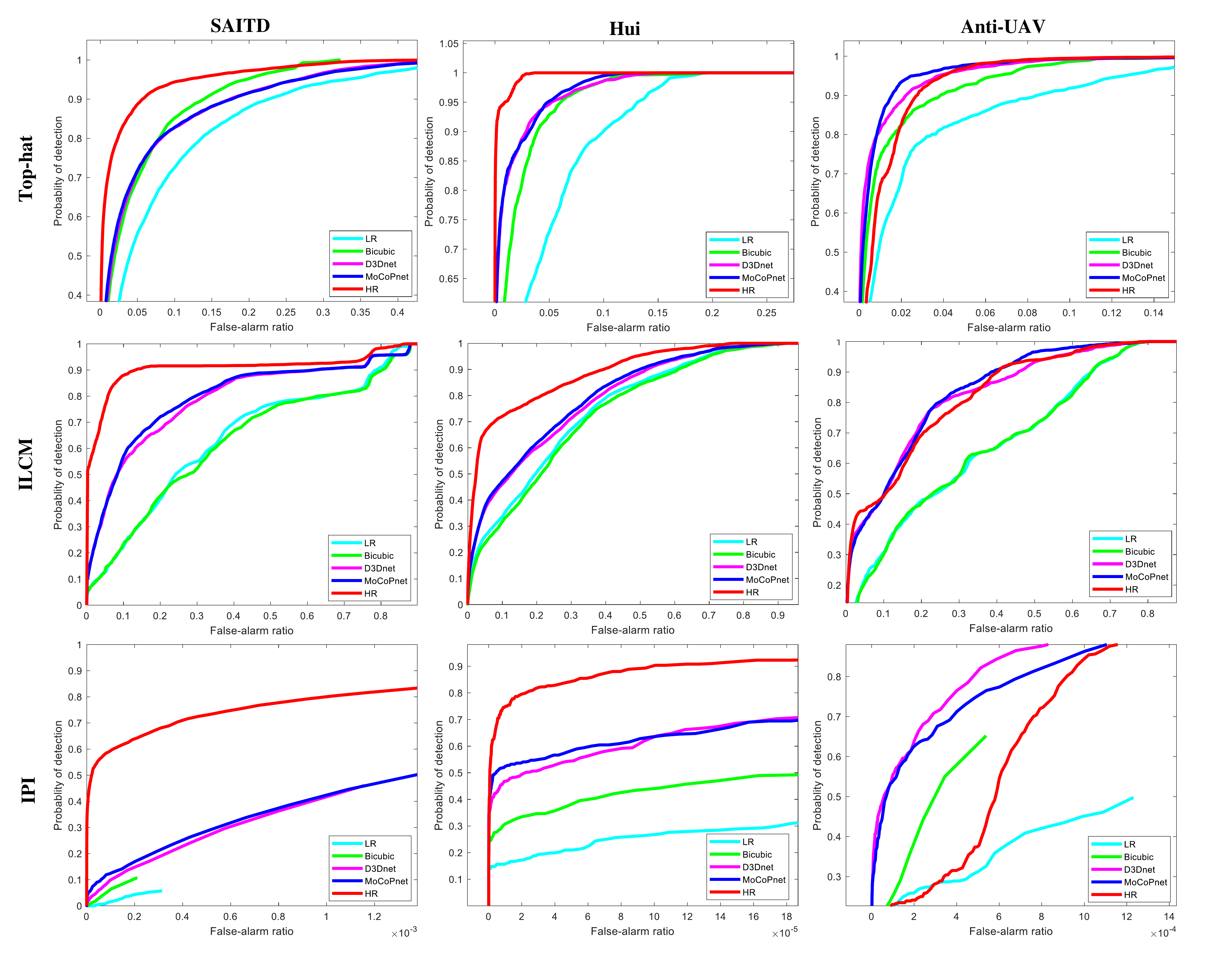}
		\caption{{ROC results of Tophat, ILCM and IPI achieved on super-resolved LR images in SAITD, Hui and Anti-UAV datasets.}}\label{low_ROC}
	\end{figure*}

	\begin{figure*}[t]
		\centering
		\includegraphics[width=18cm]{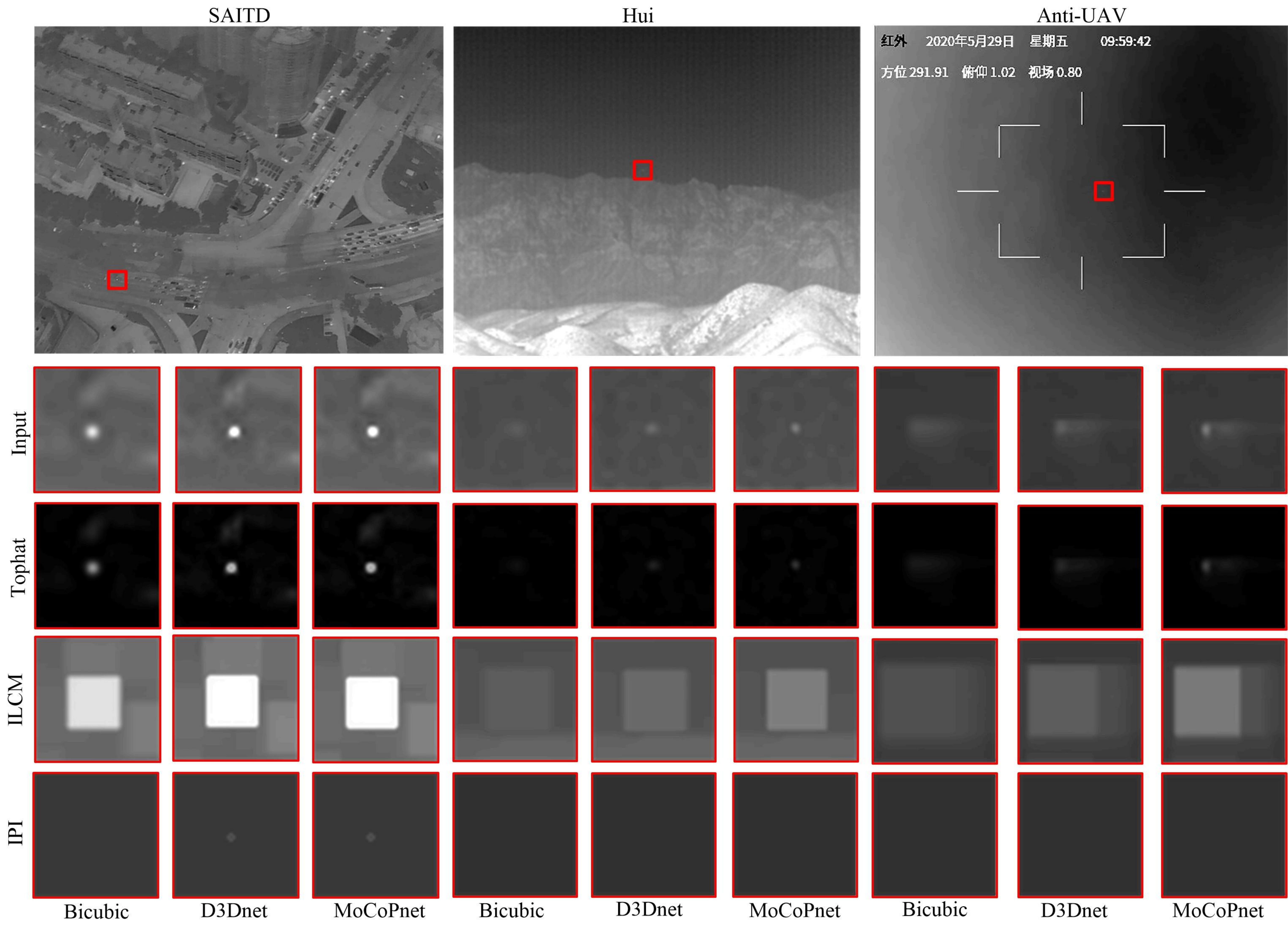}
		\caption{{Qualitative results of super-resolved HR image and detection results in SAITD, Hui and Anti-UAV datasets.}}\label{fig_21_256}%
	\end{figure*}

	\begin{figure*}[t]
	\centering
	\includegraphics[width=17cm]{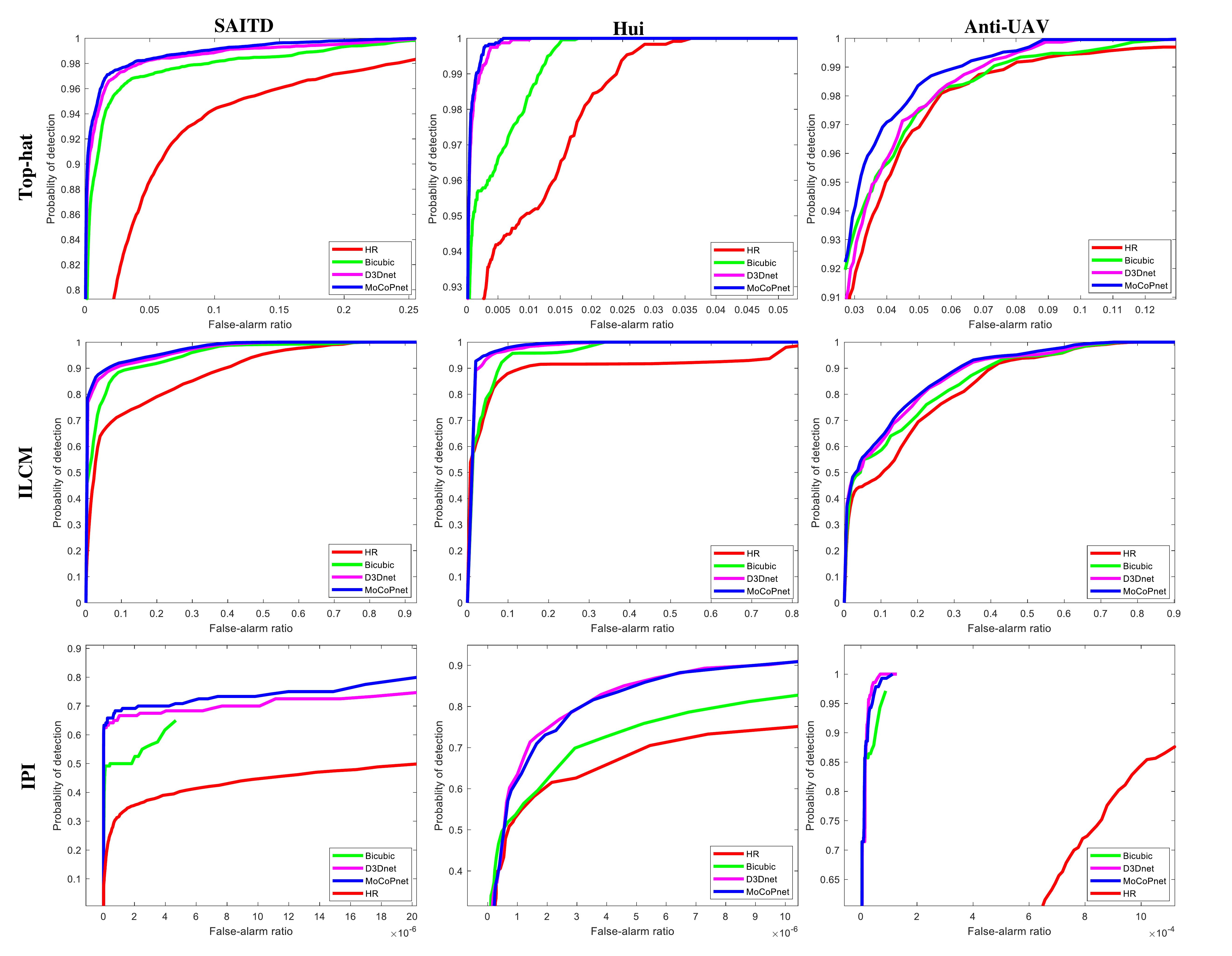}
	\caption{{ROC results of Tophat, ILCM and IPI achieved on super-resolved HR images in SAITD, Hui and Anti-UAV datasets.}}\label{fig-roc256}
\end{figure*}

\begin{figure}[t]
	\centering
	\includegraphics[width=9cm]{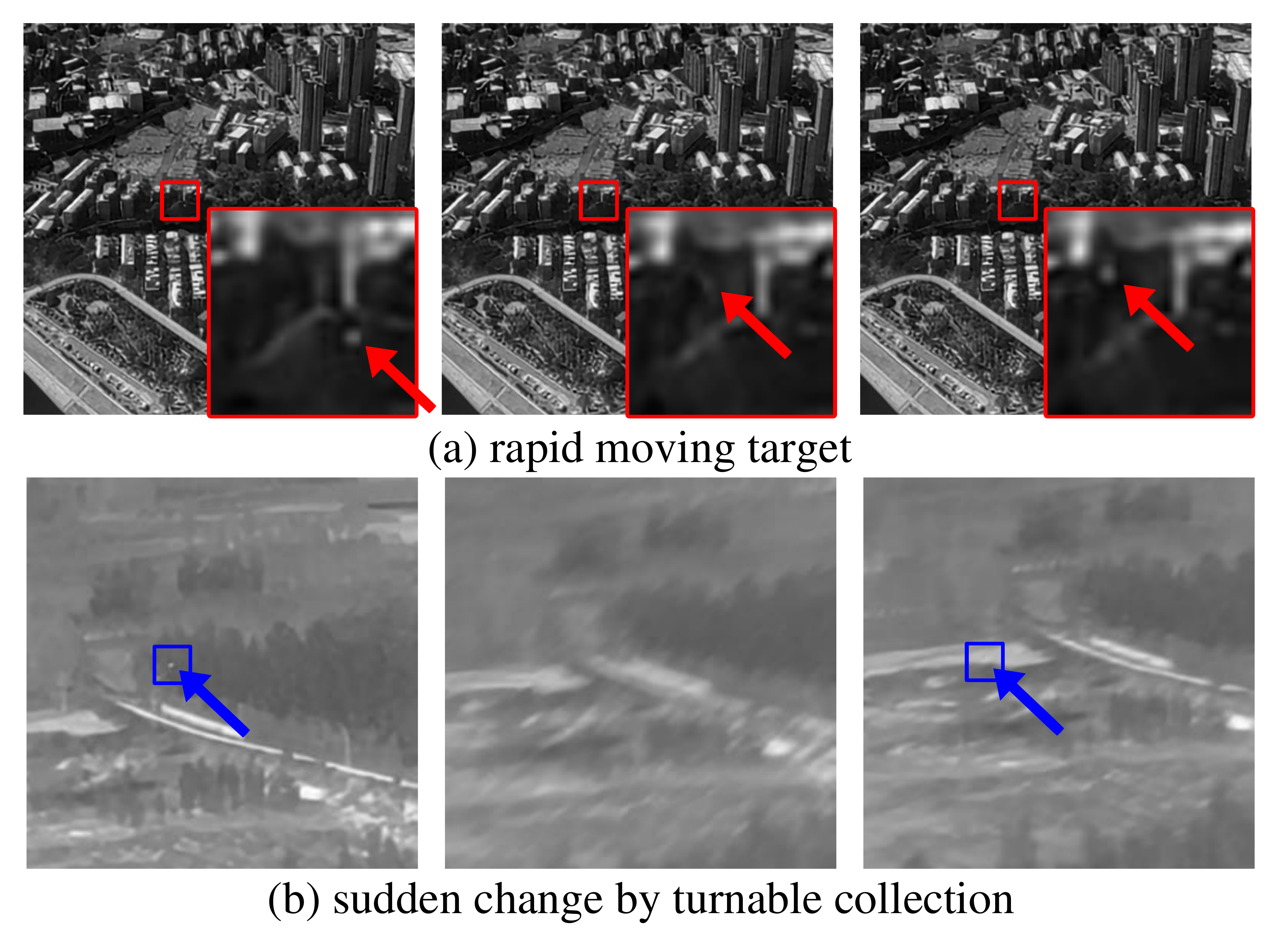}
	\caption{{Failure cases caused by (a) repaid moving targets and (b) sudden changes caused by turnable collection. Red and blue arrows indicate the locations of the targets. }}\label{failure}\vspace{-3pt}
\end{figure}

	{In this subsection, we select three typical infrared small target detection algorithms (Top-hat \cite{tophat}, ILCM\cite{ILCM}, IPI \cite{IPI}) to perform detection on super-resolved infrared images. The parameters of the three infrared small target detection algorithms are shown in Table~\ref{excel_detection}. When 4$\times$ SR is performed on HR images, the size of filters, block and stride, as well as the true detection threshold $\tau$ are enlarged by 4 times respectively. When 4$\times$ downsampling is performed on HR images, the filter sizes of Top-hat and ILCM are set to 3$\times$3. The block sizes and the stride of IPI are set to 15$\times$15 and 3. The true detection threshold $\tau$ is set to 3.0.} For simplicity, we only use the best two super-resolved results of D3Dnet and MoCoPnet to perform detection. We also introduce bicubicly upsampled (Bicubic) images and HR images as the baseline results.

	\subsubsection{Detection on Synthetic Images}\label{sec111}
	{The quantitative detection results of super-resolved LR images are listed in Table~\ref{excel_dectectLR}. It can be observed that the SNRG, SCRG and CG of the super-resolved images are generally higher than the Bicubic images. This demonstrates that SR algorithms can effectively improve the contrast between the target and the background, thus promoting the detection performance. It is worth noting that the SNRG, SCRG and CG scores of D3Dnet and MoCoPnet can even surpass those of HR. This is because, SR algorithms can perform better on the high-frequency small targets than the low-frequency local background, thus achieving improved target contrast than HR images. In addition, Bicubic can achieve the highest BSF score in most cases. This is because SR algorithms act on the entire image, which enhances targets and background simultaneously and detection algorithms have better filtering performance in smoothly changing background. Note that, BSF of MoCoPnet is generally higher than that of D3Dnet. This is because MoCoPnet can focus on recovering the local salient features in the image and further improve the contrast between targets and background, which benefits the detection performance.}

	{The qualitative results of super-resolved LR images and detection results are shown in Fig.~\ref{fig_18_256}. In the LR images, the targets intensity are very low (\textit{e.g.,} the targets in SAITD and Anti-UAV are almost invisible). In the super-resolved images, the targets intensity are higher and closer to the HR images. This is because, SR algorithms can effectively use the spatio-temporal information to enhance the target contrast. Note that, our MoCoPnet is more robust to large motion caused by turntable collections \cite{infrareddata} (\textit{i.e.,} artifacts in the zoom-in region of D3Dnet in Hui dataset). In addition, the neighborhood noise in HR image are suppressed by the way of downsampling and then super-resolution (\textit{e.g.,} point noise are not exist in the zoom-in regions of Hui and Anti-UAV datasets). Then, we perform detection on the super-resolved images. It can be observed in Fig.~\ref{fig_18_256} that all the detection algorithms have poor performance on the Bicubic images (\textit{e.g.,} the target intensity in the target image is very low and almost invisible in all detection results). This is because, bicubic interpolation cannot introduce additional information. However, the targets intensity in the target images of super-resolved images are higher than the Bicubic images. Among the super-resolved images, MoCoPnet is superior than D3Dnet in improving the target contrast due to the center-oriented gradient-aware feature extraction of CD-RG and the effective spatio-temporal information exploitation of LSTA.}

	{To evaluate the detection performance comprehensively, we further calculate the ROC results which are shown in Fig.~\ref{low_ROC}. Note that, ROC results on LR and HR image are used as the baseline results. The targets in HR images have the highest intensity. Therefore, high detection probability and low false alarm probability can be obtained and the detection probability reaches 1 faster (\textit{e.g.,} The ROC results reach 1 the fast in SAITD and Hui datasets). Downsampling leads to target intensity reduction, thus reducing the detection probability and increasing the false alarms probability. Bicubic introduces no additional image prior information, therefore, LR and Bicubic have the worst detection performance and the ROC results are significant lower than other algorithms (\textit{e.g.,} the ROC results of LR are the lowest and those of Bicubic are the second lowest except the ROC of Tophat in the SAITD dataset). SR algorithms can introduce prior information to improve the contrast between targets and background, thus achieving improved detection accuracy (\textit{e.g.,} The ROC results of MoCoPnet and D3Dnet are higher than Bicubic in SAITD and Hui datasets and even higher than HR in Anti-UAV dataset). Note that, false alarm rates of LR and Bicubic can only reach a relatively low value. This is because, IPI achieves detection by sparse and low rank recovery, which significantly decreases the false alarm rate than Tophat and ILCM. From another point, IPI suffers low detection rate of low contrast targets. Therefore, the ROC curves of Bicubic and LR images are shorter than those of HR and super-resolved images. The above experimental results show that SR algorithms can recover high-contrast targets, thus improving the detection performance.}

	\subsubsection{Detection on Real Images} {The quantitative detection results of super-resolved HR images are listed in Table~\ref{excel_dectectHR}. It can be observed that the detection performance of SR algorithms is superior to Bicubic. This demonstrates that MoCoPnet and D3Dnet can effectively improve the contrast between targets and background, resulting in performance gain of detection. Among SR algorithms, due to the superior performance of SR and target enhancement by our well-designed modules, MoCoPnet can achieve the best SNRG, SCRG and CG scores in most cases. Note that, the SNRG and SCRG scores (achieved by IPI) of MoCoPnet in Anti-UAV dataset are 7-8 orders of magnitude lower than those of Bicubic and D3Dnet. First of all, MoCoPnet can achieve highest scores of CG. This demonstrates the target intensity can be effectively and further enhanced by MoCoPnet. Then, the differences come from the performance of background suppression. Since MoCoPnet can achieve higher scores of SR performance than Bicubic and D3Dnet, the local backgrounds of Bicubic and D3Dnet are more gentle and detection algorithms can achieve better suppression performance. IPI is superior in suppressing background clutter, therefore, sometimes the local backgrounds in the target image of Bicubic and D3Dnet are zero. Since we add $\epsilon$ to each denominator in equations \ref{eq11}-\ref{eq14} to prevent it from being zero, SNRG and SCRG scores can be very large due to completely suppressed background. In addition, bicubic interpolation suppresses the high-frequency components to a certain extent, resulting in optimal BSF value.}

	{The qualitative results of super-resolved HR images and detection results are shown in Fig.~\ref{fig_21_256}. It can be observed that the targets of Bicubic images are blur while SR can enhance the intensity of target (\textit{e.g.,} the highlighted and sharpened targets). After processed by SR algorithms, we then perform detection on the super-resolved images. Note that, SR algorithms can effectively improve the intensity of targets and the contrast against background, resulting in better detection performance.}

	{To evaluate the detection performance comprehensively, we further present the ROC results in Fig.~\ref{fig-roc256}. Note that, ROC results on HR image are used as the baseline results. It can be observed that SR algorithms can improve the detection probability and reduce false alarm probability in most cases. Compared with D3Dnet, MoCoPnet can further improve the target contrast, thus promoting the detection performance. Note that, false alarm rates of Bicubic can only reach a relatively low value. This is because, IPI achieves detection by sparse and low rank recovery, which significantly decreases the false alarm rate than Tophat and ILCM. In other words, IPI suffers low detection rate of low contrast targets.}
	%\begin{figure*}[t]
	%	\centering
	%	\includegraphics[width=18.2cm]{Fig/high_detect.pdf}
	%	\caption{Qualitive results of super-resolved LR image and detection results in data18 and data20.}\label{fig_18_256}%
	%\end{figure*}

	{\subsection{Limitation}
	 The proposed method fails when the image sequence contains repaid moving targets (Fig.\ref{failure}(a)) or sudden changes (Fig.\ref{failure}(b)) caused by turnable collections. As we do not have a specific design for handling large motion and sudden change, the motion compensation by LSTAs in these cases can be wrong and our approach may not be able to effectively recover the targets. In future work, we aim to improve the robustness of our method to large motion and sudden change.}

	\section{Conclusion}\label{sec5}
	{In this paper, we propose a local motion and contrast prior driven deep network (MoCoPnet) for infrared small target super-resolution. Experimental results show that MoCoPnet can effectively recover the image details and enhance the contrast between targets and background. Based on the super-resolved images, we further investigate the effect of SR algorithms on detection performance. Experimental results show that MoCoPnet can improve the performance of infrared small target detection.}

	\bibliographystyle{IEEEtran}
	\bibliography{reference}

\begin{thebibliography}{10}
\providecommand{\url}[1]{#1}
\csname url@samestyle\endcsname
\providecommand{\newblock}{\relax}
\providecommand{\bibinfo}[2]{#2}
\providecommand{\BIBentrySTDinterwordspacing}{\spaceskip=0pt\relax}
\providecommand{\BIBentryALTinterwordstretchfactor}{4}
\providecommand{\BIBentryALTinterwordspacing}{\spaceskip=\fontdimen2\font plus
\BIBentryALTinterwordstretchfactor\fontdimen3\font minus
  \fontdimen4\font\relax}
\providecommand{\BIBforeignlanguage}[2]{{%
\expandafter\ifx\csname l@#1\endcsname\relax
\typeout{** WARNING: IEEEtran.bst: No hyphenation pattern has been}%
\typeout{** loaded for the language `#1'. Using the pattern for}%
\typeout{** the default language instead.}%
\else
\language=\csname l@#1\endcsname
\fi
#2}}
\providecommand{\BIBdecl}{\relax}
\BIBdecl

\bibitem{ACM}
Y.~Dai, Y.~Wu, F.~Zhou, and K.~Barnard, ``Asymmetric contextual modulation for
  infrared small target detection,'' in \emph{Proceedings of the IEEE Winter
  Conference on Applications of Computer Vision}, 2021.

\bibitem{liu2020small}
H.-K. Liu, L.~Zhang, and H.~Huang, ``Small target detection in infrared videos
  based on spatio-temporal tensor model,'' \emph{IEEE Transactions on
  Geoscience and Remote Sensing}, vol.~58, no.~12, pp. 8689--8700, 2020.

\bibitem{sun2020infrared}
Y.~Sun, J.~Yang, and W.~An, ``Infrared dim and small target detection via
  multiple subspace learning and spatial-temporal patch-tensor model,''
  \emph{IEEE Transactions on Geoscience and Remote Sensing}, vol.~59, no.~5,
  pp. 3737--3752, 2020.

\bibitem{detect61}
Y.~Qin and B.~Li, ``Effective infrared small target detection utilizing a novel
  local contrast method,'' \emph{IEEE Geoscience and Remote Sensing Letters},
  vol.~13, no.~12, pp. 1890--1894, 2016.

\bibitem{ALCNet}
Y.~{Dai}, Y.~{Wu}, F.~{Zhou}, and K.~{Barnard}, ``Attentional local contrast
  networks for infrared small target detection,'' \emph{IEEE Transactions on
  Geoscience and Remote Sensing}, pp. 1--12, 2021.

\bibitem{survey12}
M.~Irani and S.~Peleg, ``Improving resolution by image registration,''
  \emph{Graphical Models and Image Processing}, vol.~53, no.~3, pp. 231--239,
  1991.

\bibitem{survey10}
R.~Keys, ``Cubic convolution interpolation for digital image processing,''
  \emph{IEEE Transactions on Acoustics, Speech, and Signal Processing},
  vol.~29, no.~6, pp. 1153--1160, 1981.

\bibitem{survey14}
J.~Sun, Z.~Xu, and H.~Shum, ``Image super-resolution using gradient profile
  prior,'' in \emph{Proceedings of the IEEE Conference on Computer Vision and
  Pattern Recognition}.\hskip 1em plus 0.5em minus 0.4em\relax IEEE, 2008, pp.
  1--8.

\bibitem{survey13}
G.~Freedman and R.~Fattal, ``Image and video upscaling from local
  self-examples,'' \emph{ACM Transactions on Graphics}, vol.~30, no.~2, pp.
  1--11, 2011.

\bibitem{survey15}
K.~Kim and Y.~Kwon, ``Single-image super-resolution using sparse regression and
  natural image prior,'' \emph{IEEE Transactions on Pattern Analysis and
  Machine Intelligence}, vol.~32, no.~6, pp. 1127--1133, 2010.

\bibitem{survey16}
Z.~Xiong, X.~Sun, and F.~Wu, ``Robust web image/video super-resolution,''
  \emph{IEEE Transactions on Image Processing}, vol.~19, no.~8, pp. 2017--2028,
  2010.

\bibitem{survey18}
H.~Chang, D.~Yeung, and Y.~Xiong, ``Super-resolution through neighbor
  embedding,'' in \emph{Proceedings of the IEEE Computer Society Conference on
  Computer Vision and Pattern Recognition}, vol.~1.\hskip 1em plus 0.5em minus
  0.4em\relax IEEE, 2004, pp. I--I.

\bibitem{survey20}
J.~Yang, J.~Wright, T.~Huang, and Y.~Ma, ``Image super-resolution as sparse
  representation of raw image patches,'' in \emph{Proceedings of the IEEE
  Conference on Computer Vision and Pattern Recognition}.\hskip 1em plus 0.5em
  minus 0.4em\relax IEEE, 2008, pp. 1--8.

\bibitem{survey21}
{Yang, Jianchao and Wright, John and Huang, Thomas and Ma, Yi}, ``Image
  super-resolution via sparse representation,'' \emph{IEEE Transactions on
  Image Processing}, vol.~19, pp. 2861--2873, 2010.

\bibitem{RCAN}
Y.~Zhang, K.~Li, K.~Li, L.~Wang, B.~Zhong, and Y.~Fu, ``Image super-resolution
  using very deep residual channel attention networks,'' in \emph{Proceedings
  of the European Conference on Computer Vision}, 2018, pp. 286--301.

\bibitem{PAM}
L.~Wang, Y.~Guo, Y.~Wang, Z.~Liang, Z.~Lin, J.~Yang, and W.~An, ``Parallax
  attention for unsupervised stereo correspondence learning,'' \emph{IEEE
  Transactions on Pattern Analysis and Machine Intelligence}, 2020.

\bibitem{SRCNN}
C.~Dong, C.~Loy, K.~He, and X.~Tang, ``Learning a deep convolutional network
  for image super-resolution,'' in \emph{Proceedings of the European Conference
  on Computer Vision}, 2014, pp. 184--199.

\bibitem{VDSR}
J.~Kim, L.~Kwon, and L.~Mu, ``Accurate image super-resolution using very deep
  convolutional networks,'' in \emph{Proceedings of the IEEE conference on
  Computer Vision and Pattern Recognition}, 2016, pp. 1646--1654.

\bibitem{EDSR}
B.~Lim, S.~Son, H.~Kim, S.~Nah, and K.~Mu~Lee, ``Enhanced deep residual
  networks for single image super-resolution,'' in \emph{Proceedings of the
  IEEE Conference on Computer Vision and Pattern Recognition Workshops}, 2017,
  pp. 136--144.

\bibitem{li2019lightweight}
J.~Li, Y.~Yuan, K.~Mei, and F.~Fang, ``Lightweight and accurate recursive
  fractal network for image super-resolution,'' in \emph{Proceedings of the
  IEEE/CVF International Conference on Computer Vision Workshops}, 2019, pp.
  0--0.

\bibitem{DRCN}
J.~Kim, L.~Kwon, and L.~Mu, ``Deeply-recursive convolutional network for image
  super-resolution,'' in \emph{Proceedings of the IEEE Conference on Computer
  Vision and Pattern Recognition}, 2016, pp. 1637--1645.

\bibitem{DRRN}
Y.~Tai, J.~Yang, and X.~Liu, ``Image super-resolution via deep recursive
  residual network,'' in \emph{Proceedings of the IEEE Conference on Computer
  Vision and Pattern Recognition}, 2017, pp. 3147--3155.

\bibitem{PR1}
K.~Jiang, Z.~Wang, P.~Yi, and J.~Jiang, ``Hierarchical dense recursive network
  for image super-resolution,'' \emph{Pattern Recognition}, vol. 107, p.
  107475, 2020.

\bibitem{li2020mdcn}
J.~Li, F.~Fang, J.~Li, K.~Mei, and G.~Zhang, ``{MDCN}: Multi-scale dense cross
  network for image super-resolution,'' \emph{IEEE Transactions on Circuits and
  Systems for Video Technology}, vol.~31, no.~7, pp. 2547--2561, 2020.

\bibitem{SRDenseNet}
T.~Tong, G.~Li, X.~Liu, and Q.~Gao, ``Image super-resolution using dense skip
  connections,'' in \emph{Proceedings of the IEEE conference on Computer Vision
  and Pattern Recognition}, 2017, pp. 4799--4807.

\bibitem{RDN}
Y.~Zhang, Y.~Tian, Y.~Kong, B.~Zhong, and Y.~Fu, ``Residual dense network for
  image super-resolution,'' in \emph{Proceedings of the IEEE Conference on
  Computer Vision and Pattern Recognition}, 2018, pp. 2472--2481.

\bibitem{SAN}
T.~Dai, J.~Cai, Y.~Zhang, S.-T. Xia, and L.~Zhang, ``Second-order attention
  network for single image super-resolution,'' in \emph{Proceedings of the IEEE
  Conference on Computer Vision and Pattern Recognition}, 2019, pp.
  11\,065--11\,074.

\bibitem{SRGAN}
C.~Ledig, L.~Theis, F.~Husz{\'a}r, J.~Caballero, A.~Cunningham, A.~Acosta,
  A.~Aitken, A.~Tejani, J.~Totz, Z.~Wang \emph{et~al.}, ``Photo-realistic
  single image super-resolution using a generative adversarial network,'' in
  \emph{Proceedings of the IEEE Conference on Computer Vision and Pattern
  Recognition}, 2017, pp. 4681--4690.

\bibitem{ESRGAN}
X.~Wang, K.~Yu, S.~Wu, J.~Gu, Y.~Liu, C.~Dong, Y.~Qiao, and C.~Change~Loy,
  ``Esrgan: Enhanced super-resolution generative adversarial networks,'' in
  \emph{Proceedings of the European Conference on Computer Vision}, 2018, pp.
  0--0.

\bibitem{D3Dnet}
X.~Ying, L.~Wang, Y.~Wang, W.~Sheng, W.~An, and Y.~Guo, ``Deformable 3d
  convolution for video super-resolution,'' \emph{IEEE Signal Processing
  Letters}, vol.~27, pp. 1500--1504, 2020.

\bibitem{videoSR-t1}
C.~Liu and D.~Sun, ``On bayesian adaptive video super resolution,'' \emph{IEEE
  Transactions on Pattern Analysis and Machine Intelligence}, vol.~36, no.~2,
  pp. 346--360, 2013.

\bibitem{videoSR-t2}
M.~Ben-Ezra, A.~Zomet, and S.~K. Nayar, ``Video super-resolution using
  controlled subpixel detector shifts,'' \emph{IEEE Transactions on Pattern
  Analysis and Machine Intelligence}, vol.~27, no.~6, pp. 977--987, 2005.

\bibitem{video-m1}
R.~Liao, X.~Tao, R.~Li, Z.~Ma, and J.~Jia, ``Video super-resolution via deep
  draft-ensemble learning,'' in \emph{Proceedings of the IEEE International
  Conference on Computer Vision}, 2015, pp. 531--539.

\bibitem{VESPCN}
J.~Caballero, C.~Ledig, A.~Aitken, A.~Acosta, J.~Totz, Z.~Wang, and W.~Shi,
  ``Real-time video super-resolution with spatio-temporal networks and motion
  compensation,'' in \emph{Proceedings of the IEEE Conference on Computer
  Vision and Pattern Recognition}, 2017, pp. 4778--4787.

\bibitem{SOFVSR20}
L.~Wang, Y.~Guo, L.~Liu, Z.~Lin, X.~Deng, and W.~An, ``Deep video
  super-resolution using {HR} optical flow estimation,'' \emph{IEEE
  Transactions on Image Processing}, 2020.

\bibitem{vague2}
T.~Isobe, S.~Li, X.~Jia, S.~Yuan, G.~Slabaugh, C.~Xu, Y.-L. Li, S.~Wang, and
  Q.~Tian, ``Video super-resolution with temporal group attention,'' in
  \emph{Proceedings of the IEEE Conference on Computer Vision and Pattern
  Recognition}, 2020, pp. 8008--8017.

\bibitem{DCN1}
J.~Dai, H.~Qi, Y.~Xiong, Y.~Li, G.~Zhang, H.~Hu, and Y.~Wei, ``Deformable
  convolutional networks,'' in \emph{Proceedings of the IEEE International
  Conference on Computer Vision}, 2017, pp. 764--773.

\bibitem{DCN2}
X.~Zhu, H.~Hu, S.~Lin, and J.~Dai, ``Deformable convnets v2: More deformable,
  better results,'' in \emph{Proceedings of the IEEE Conference on Computer
  Vision and Pattern Recognition}, 2019, pp. 9308--9316.

\bibitem{TDAN}
Y.~Tian, Y.~Zhang, Y.~Fu, and C.~Xu, ``{TDAN}: Temporally deformable alignment
  network for video super-sesolution,'' in \emph{Proceedings of the IEEE
  International Conference on Computer Vision and Pattern Recognition}, 2020.

\bibitem{EDVR}
X.~Wang, K.~C. Chan, K.~Yu, C.~Dong, and C.~Change~Loy, ``{EDVR}: Video
  restoration with enhanced deformable convolutional networks,'' in
  \emph{Proceedings of the IEEE Conference on Computer Vision and Pattern
  Recognition Workshops}, 2019, pp. 0--0.

\bibitem{DUF-VSR}
Y.~Jo, S.~Wug~Oh, J.~Kang, and S.~Joo~Kim, ``Deep video super-resolution
  network using dynamic upsampling filters without explicit motion
  compensation,'' in \emph{Proceedings of the IEEE Conference on Computer
  Vision and Pattern Recognition}, 2018, pp. 3224--3232.

\bibitem{FSTRN}
S.~Li, F.~He, B.~Du, L.~Zhang, Y.~Xu, and D.~Tao, ``Fast spatio-temporal
  residual network for video super-resolution,'' in \emph{Proceedings of the
  IEEE Conference on Computer Vision and Pattern Recognition}, 2019.

\bibitem{over-video30}
Y.~Huang, W.~Wang, and L.~Wang, ``Video super-resolution via bidirectional
  recurrent convolutional networks,'' \emph{IEEE Transactions on Pattern
  Analysis and Machine Intelligence}, vol.~40, no.~4, pp. 1015--1028, 2017.

\bibitem{over-video35}
X.~Zhu, Z.~Li, X.~Zhang, C.~Li, Y.~Liu, and Z.~Xue, ``Residual invertible
  spatio-temporal network for video super-resolution,'' in \emph{Proceedings of
  the AAAI Conference on Artificial Intelligence}, vol.~33, 2019, pp.
  5981--5988.

\bibitem{PFNL}
P.~Yi, Z.~Wang, K.~Jiang, J.~Jiang, and J.~Ma, ``Progressive fusion video
  super-resolution network via exploiting non-local spatio-temporal
  correlations,'' in \emph{Proceedings of the IEEE International Conference on
  Computer Vision}, 2019, pp. 3106--3115.

\bibitem{InMao}
Y.~Mao, Y.~Wang, J.~Zhou, and H.~Jia, ``An infrared image super-resolution
  reconstruction method based on compressive sensing,'' \emph{Infrared Physics
  \& Technology}, vol.~76, pp. 735--739, 2016.

\bibitem{InZhang}
X.~Zhang, C.~Li, Q.~Meng, S.~Liu, Y.~Zhang, and J.~Wang, ``Infrared image super
  resolution by combining compressive sensing and deep learning,''
  \emph{Sensors}, vol.~18, no.~8, p. 2587, 2018.

\bibitem{InHan}
T.~Y. Han, Y.~J. Kim, and B.~C. Song, ``Convolutional neural network-based
  infrared image super resolution under low light environment,'' in
  \emph{Proceedings of the European Signal Processing Conference}.\hskip 1em
  plus 0.5em minus 0.4em\relax IEEE, 2017, pp. 803--807.

\bibitem{InHe}
Z.~He, S.~Tang, J.~Yang, Y.~Cao, M.~Y. Yang, and Y.~Cao, ``Cascaded deep
  networks with multiple receptive fields for infrared image
  super-resolution,'' \emph{IEEE transactions on circuits and systems for video
  technology}, vol.~29, no.~8, pp. 2310--2322, 2018.

\bibitem{InLiu}
Q.-M. Liu, R.-S. Jia, Y.-B. Liu, H.-B. Sun, J.-Z. Yu, and H.-M. Sun, ``Infrared
  image super-resolution reconstruction by using generative adversarial network
  with an attention mechanism,'' \emph{Applied Intelligence}, vol.~51, no.~4,
  pp. 2018--2030, 2021.

\bibitem{IERN}
L.~Chen, R.~Tang, M.~Anisetti, and X.~Yang, ``A lightweight iterative error
  reconstruction network for infrared image super-resolution in smart grid,''
  \emph{Sustainable Cities and Society}, vol.~66, p. 102520, 2021.

\bibitem{PSRGAN}
Y.~Huang, Z.~Jiang, R.~Lan, S.~Zhang, and K.~Pi, ``Infrared image
  super-resolution via transfer learning and psrgan,'' \emph{IEEE Signal
  Processing Letters}, vol.~28, pp. 982--986, 2021.

\bibitem{ChaSNet}
K.~Prajapati, V.~Chudasama, H.~Patel, A.~Sarvaiya, K.~P. Upla, K.~Raja,
  R.~Ramachandra, and C.~Busch, ``Channel split convolutional neural network
  (chasnet) for thermal image super-resolution,'' in \emph{Proceedings of the
  IEEE Conference on Computer Vision and Pattern Recognition}, 2021, pp.
  4368--4377.

\bibitem{Visible-Assisted}
X.~Yang, M.~Zhang, W.~Li, and R.~Tao, ``Visible-assisted infrared image
  super-resolution based on spatial attention residual network,'' \emph{IEEE
  Geoscience and Remote Sensing Letters}, vol.~19, pp. 1--5, 2021.

\bibitem{SENet}
J.~Hu, L.~Shen, and G.~Sun, ``Squeeze-and-excitation networks,'' in
  \emph{Proceedings of the IEEE conference on computer vision and pattern
  recognition}, 2018, pp. 7132--7141.

\bibitem{FeatureA}
S.~Anwar and N.~Barnes, ``Real image denoising with feature attention,'' in
  \emph{Proceedings of the IEEE International Conference on Computer Vision},
  2019, pp. 3155--3164.

\bibitem{Wang2021Exploring}
L.~Wang, X.~Dong, Y.~Wang, X.~Ying, Z.~Lin, W.~An, and Y.~Guo, ``Exploring
  sparsity in image super-resolution for efficient inference,'' in
  \emph{Proceedings of the IEEE conference on Computer Vision and Pattern
  Recognition}, 2021.

\bibitem{IPI}
C.~Gao, D.~Meng, Y.~Yang, Y.~Wang, X.~Zhou, and A.~G. Hauptmann, ``Infrared
  patch-image model for small target detection in a single image,'' \emph{IEEE
  Transactions on Image Processing}, vol.~22, no.~12, pp. 4996--5009, 2013.

\bibitem{detect59}
C.~P. Chen, H.~Li, Y.~Wei, T.~Xia, and Y.~Y. Tang, ``A local contrast method
  for small infrared target detection,'' \emph{IEEE Transactions on Geoscience
  and Remote Sensing}, vol.~52, no.~1, pp. 574--581, 2013.

\bibitem{liu2021non}
T.~Liu, J.~Yang, B.~Li, C.~Xiao, Y.~Sun, Y.~Wang, and W.~An, ``Non-convex
  tensor low-rank approximation for infrared small target detection,''
  \emph{IEEE Transactions on Geoscience and Remote Sensing}, 2021.

\bibitem{DNAnet}
B.~Li, C.~Xiao, L.~Wang, Y.~Wang, Z.~Lin, M.~Li, W.~An, and Y.~Guo, ``Dense
  nested attention network for infrared small target detection,'' \emph{arXiv
  preprint arXiv:2106.00487}, 2021.

\bibitem{application}
I.~Reed, R.~Gagliardi, and H.~Shao, ``Application of three-dimensional
  filtering to moving target detection,'' \emph{IEEE Transactions on Aerospace
  and Electronic Systems}, no.~6, pp. 898--905, 1983.

\bibitem{Novel}
N.~J. Gordon, D.~J. Salmond, and A.~F. Smith, ``Novel approach to
  nonlinear/non-gaussian bayesian state estimation,'' \emph{IEE proceedings
  Part F - Radar and Signal Processing}, vol. 140, no.~2, pp. 107--113, 1993.

\bibitem{improved}
Z.~Chen, M.~Tian, Y.~Bo, and X.~Ling, ``Improved infrared small target
  detection and tracking method based on new intelligence particle filter,''
  \emph{Computational Intelligence}, vol.~34, no.~3, pp. 917--938, 2018.

\bibitem{CDC}
Z.~Yu, C.~Zhao, Z.~Wang, Y.~Qin, Z.~Su, X.~Li, F.~Zhou, and G.~Zhao,
  ``Searching central difference convolutional networks for face
  anti-spoofing,'' in \emph{Proceedings of the IEEE Conference on Computer
  Vision and Pattern Recognition}, 2020, pp. 5295--5305.

\bibitem{CDC2}
Z.~Su, W.~Liu, Z.~Yu, D.~Hu, Q.~Liao, Q.~Tian, M.~Pietikäinen, and L.~Liu,
  ``Pixel difference networks for efficient edge detection,'' in
  \emph{Proceedings of the IEEE International Conference on Computer Vision},
  2021.

\bibitem{Wang_2021_CVPR}
Y.~Wang, X.~Ying, L.~Wang, J.~Yang, W.~An, and Y.~Guo, ``Symmetric parallax
  attention for stereo image super-resolution,'' in \emph{Proceedings of the
  IEEE Conference on Computer Vision and Pattern Recognition Workshops}, June
  2021, pp. 766--775.

\bibitem{DSFNet}
C.~Xiao, Q.~Yin, X.~Ying, R.~Li, S.~Wu, M.~Li, L.~Liu, W.~An, and Z.~Chen,
  ``Dsfnet: Dynamic and static fusion network for moving object detection in
  satellite videos,'' \emph{IEEE Geoscience and Remote Sensing Letters},
  vol.~19, pp. 1--5, 2021.

\bibitem{RDB}
Y.~Zhang, Y.~Tian, Y.~Kong, B.~Zhong, and Y.~Fu, ``Residual dense network for
  image super-resolution,'' in \emph{Proceedings of the IEEE conference on
  Computer Vision and Pattern Recognition}, 2018, pp. 2472--2481.

\bibitem{infrareddata}
B.~Hui, Z.~Song, H.~Fan, P.~Zhong, W.~Hu, X.~Zhang, J.~Ling, H.~Su, W.~Jin,
  Y.~Zhang, and Y.~Bai, ``A dataset for infrared detection and tracking of
  dim-small aircraft targets under ground / air background,'' \emph{China
  Scientific Data}, 2020.

\bibitem{SAITD}
X.~Sun, L.~Guo, W.~Zhang, Z.~Wang, Y.~Hou, Z.~Li, and X.~Teng, ``A dataset for
  small infrared moving target detection under clutter background,''
  \emph{Chinese Scientific Data}, 2021.

\bibitem{Anti-UAV}
J.~Zhao, G.~Wang, J.~Li, L.~Jin, N.~Fan, M.~Wang, X.~Wang, T.~Yong, Y.~Deng,
  Y.~Guo, and S.~Ge, ``The 2nd anti-uav workshop \& challenge,''
  \url{https://anti-uav.github.io/}, 2021.

\bibitem{Adam}
D.~P. Kingma and J.~Ba, ``Adam: A method for stochastic optimization,'' in
  \emph{Proceedings of the International Conference on Learning
  Representations}, 2015.

\bibitem{visual111}
N.~Komodakis and S.~Zagoruyko, ``Paying more attention to attention: improving
  the performance of convolutional neural networks via attention transfer,'' in
  \emph{International Conference on Learning Representations}, 2017.

\bibitem{VSRnet}
A.~Kappeler, S.~Yoo, Q.~Dai, and A.~K. Katsaggelos, ``Video super-resolution
  with convolutional neural networks,'' \emph{IEEE Transactions on
  Computational Imaging}, vol.~2, no.~2, pp. 109--122, 2016.

\bibitem{tophat}
J.~Rivest and R.~Fortin, ``Detection of dim targets in digital infrared imagery
  by morphological image processing,'' \emph{Optical Engineering}, vol.~35,
  no.~7, pp. 1886--1893, 1996.

\bibitem{ILCM}
X.~Zhang, Q.~Ding, H.~Luo, H.~Bin, C.~Zheng, and Z.~Junchao, ``Infrared dim
  target detection algorithm based on improved lcm,'' \emph{Infrared and Laser
  Engineering}, vol.~46, no.~7, pp. 726\,002--0\,726\,002, 2017.

\bibitem{vague1}
Y.~Lu, J.~Valmadre, H.~Wang, J.~Kannala, M.~Harandi, and P.~Torr, ``Devon:
  Deformable volume network for learning optical flow,'' in \emph{Proceedings
  of the IEEE Winter Conference on Applications of Computer Vision}, 2020, pp.
  2705--2713.

\end{thebibliography}

\end{document}